\documentclass[iop,apj]{emulateapj}
\usepackage{graphicx}
\usepackage{amssymb}
\usepackage{amsmath,mathrsfs}
\usepackage{url}
\usepackage{subeqnarray}
\usepackage{cases}
\usepackage{diagbox}

\shorttitle{NDAF with inner boundary torque}
\shortauthors{Xie, Lei \& Wang}

\begin{document}

\def \GRB {GRB~110918A}

\def \KW  {Konus-\textit{WIND}}
\def \SW  {\textit{Swift}}
\def \BAT {\textit{Swift}-BAT}
\newcommand{\XRT}{\textit{Swift}-XRT }
\newcommand{\UVOT}{\textit{Swift}-UVOT }
\newcommand{\INT}{\textit{INTEGRAL} }
\newcommand{\SPIACS}{SPI-ACS }
\newcommand{\MO}{\textit{Mars~Odyssey} }
\newcommand{\MESS}{\textit{MESSENGER} }
\newcommand{\NHunits}{\mbox{$10^{21}~{\rm cm}^{-2}$}}
\newcommand{\flux}{erg cm$^{-2}$ s$^{-1}$}
\newcommand{\tabincell}[2]{\begin{tabular}{@{}#1@{}}#2\end{tabular}}
%\slugcomment{Submitted to ApJ 0000-00-00, revised 0000-00-00, accepted 0000-00-00}

\title{Numerical and analytical solutions of Neutrino-Dominated Accretion Flows with a Non-Zero Torque Boundary Condition and its applications in Gamma-ray Bursts}

%%%%%%%%%%%%%%%%%%%%%%%%%%%%%%%%%%%%%%%%%%%%%%%%%%%%%%%%%%%%%%%%%%%%%%%%%%%%%%%%%%%%%%%%%%%
% AUTHOR
%%%%%%%%%%%%%%%%%%%%%%%%%%%%%%%%%%%%%%%%%%%%%%%%%%%%%%%%%%%%%%%%%%%%%%%%%%%%%%%%%%%%%%%%%%%
\author{Wei Xie$^{1}$,  Wei-Hua Lei*$^{1}$, Ding-Xiong Wang$^{1}$}
\affil{$^{1}$School of Physics, Huazhong University of Science and Technology, Wuhan 430074, China. Email: leiwh@hust.edu.cn
}


%%%%%%%%%%%%%%%%%%%%%%%%%%%%%%%%%%%%%%%%%%%%%%%%%%%%%%%%%%%%%%%%%%%%%%%%%%%%%%%%%%%%%%%%%%%
% ABSTRACT
%%%%%%%%%%%%%%%%%%%%%%%%%%%%%%%%%%%%%%%%%%%%%%%%%%%%%%%%%%%%%%%%%%%%%%%%%%%%%%%%%%%%%%%%%%%
\newpage
\begin{abstract}
A stellar mass black hole (BH) surrounded by a neutrino-dominated accretion flow (NDAF) has been discussed in a number of works as the central engine of gamma-ray bursts (GRBs). It is widely believed that NDAF cannot liberate enough energy for bright GRBs. However, these works have been based on the assumption of ``no torque" boundary condition, which is invalid when the disk is magnetized. In this paper, we present both numerical and analytical solutions for NDAFs with non-zero boundary stresses, and reexamine their properties. We find that NDAF with such boundary torque can be powerful enough to account for those bright short GRBs, energetic long GRBs and ultra-long GRBs. The disk becomes viscously unstable, which makes it possible to interpret the variability of GRB prompt emission and the steep decay phase in the early X-ray afterglow. Finally, we study the gravitational waves radiated from a processing BH-NDAF. We find that the effects of the boundary torque on the strength of the gravitational waves can be ignored.
\end{abstract}

%%%%%%%%%%%%%%%%%%%%%%%%%%%%%%%%%%%%%%%%%%%%%%%%%%%%%%%%%%%%%%%%%%%%%%%%%%%%%%%%%%%%%%%%%%%
% INTRODUCTION
%%%%%%%%%%%%%%%%%%%%%%%%%%%%%%%%%%%%%%%%%%%%%%%%%%%%%%%%%%%%%%%%%%%%%%%%%%%%%%%%%%%%%%%%%%%
\newpage

\section{Introduction}
\label{sect:intro}
The leading model of Gamma-ray burst (GRB) central engine is a hyper-accreting stellar-mass black hole (BH).
The typical accretion rate is extremely high (e.g., $0.01 - 1 M_\sun s^{-1}$), leading to a much dense and hot flow. Under such condition, photons become trapped and are inefficient in cooling the disk. The gravitational energy in the accretion flow is mainly carried by neutrino and anti-neutrinos, which annihilate and power GRB jets. These disks are therefore named ``neutrino-cooling-dominated accretion flows'', or NDAFs (e.g., \citealt[hereafter PWF99]{P99}; \citealt{KM02}).

The NDAF has been extensively investigated and usually compared with the magnetic mechanism (e.g.,\citealt[hereafter NPK01]{N01}; \citealt{KM02}; \citealt[hereafter DPN02]{D02}; \citealt{CB07,J04,J07,J10,G06,Liu07,Lei08,Lei09,Lei13a}). It was for a long time considered as an inefficient model for GRBs. This conclusion was first made by Popham et al. (1999), and enhanced by Di Matteo et al. (2002). Fan et al. (2005) found that NDAF model was disfavour in explaining the X-ray flares of GRB afterglows. Recently, detailed studies by Liu et al. (2015) show that some bright short GRBs (SGRBs) are hard to be explained with NDAF. More recently, Song et al. (2016) argued that NDAF may not be the central engine for some extremely high energy long GRBs (LGRBs).

It is worth pointing out that these works are based on the assumption of zero-torque at the inner edge of the accretion disk. This condition has been argued based on the fact that small amount of mass in the plunging region could hardly be expected to exert a force on the far heavier disk proper, or rapidly becomes causally disconnected from the disk (\citealt[hereafter NT73]{NT73}). However, as recognized by \citet[hereafter PT74]{PT74}, neither of these arguments applies to magnetic stress. This issue has become increasingly important with the realization that angular momentum transport in disk is entirely due to turbulence generated via the magnetorotational instability (MRI) (\citealt{BH91}). \citet{K99} and \citet{G99} argued that the dominant role of this magnetic stresses in angular momentum transport in the disk body should actually lead to stresses near the marginally stable orbit. Based on these considerations, \citet{AK00} studied a relativistic thin disk with non-zero torque at its inner edge. As a consequence, the additional magnetic stresses have strong effects that change the fundamental properties of the accretion flow. A more complete magnetohydrodynamical (MHD) model of a magnetized thin disk has been developed by \citet{G99}, a numerical MHD simulation has been carried out by \citet{RA01} using ZEUS code \citep{SN92}, and they verified the existence of torque at marginally stable radius due to the coupling of the plunging region to the disk through magnetic fields. The accretion of a magnetized torus in Kerr metric has been studied by \citet{G03} and \citet{D03} by using general relativistic magnetohydrodynamical (GRMHD) codes. It is found that the disk would be significantly altered by the additional stress, with wide-ranging observational consequences (\citealt{AK00}; Zimmerman et al. 2005). Some authors suggested that the episodic jets in GRBs could be reproduced by magnetized NDAF (e.g. \citealt{YZ12, Cao14}). The magnetic energy within an episodic jet is possibly dissipated via internal-collision-induced magnetic reconnection and turbulence (ICMART, \citealt{ZY11}). These works motive us to investigate the NDAF with boundary stresses. We refer this model as non-zero torque NDAF (nztNDAF), and the previous NDAF model with zero boundary torque as NDAF.

This paper is organized as follows: In section 2, we describe the NDAF model with boundary stress and general relativistic corrections. A free parameter $\eta$ is introduced to account for the magnitude of the unknown stress at the inner edge of nztNDAF. In section 3, we study the properties of the disk by solving the set of equations. Based on the solutions, we investigate the stability and total neutrino annihilation luminosity of nztNDAF. In section 4, we apply the nztNDAF model to GRBs. In section 5, we find that the variability of the GRB prompt emissions and the steep decay phase in the early X-ray afterglow can be well explained by the viscous instability in nztNDAF. In section 6, we investigate the effect of inner boundary torque on the gravitational waves radiated from a processing disk. We summarize and discuss the results of this work in section 7.
%***************************************************************

\section{NDAF with boundary stress}
\label{sect:main}
As argued by \citet{K99}, a sizeable torque would be exerted on the disk's inner edge if matter inside the marginally stable orbit $r_{\rm{ms}}$(NT73) remains magnetically connected to the disk. Hereafter, the subscript ``ms'' indicates the quantity at the marginally stable orbit. There is no characteristic or ``nature'' magnitude that one can select for the torque. In principal, additional local dissipation must accompany the additional boundary torque. In the context of a standard thin disk (SSD), such extra dissipation leads to an increment in disk radiation. For this reason, in \citet{AK00}, the significance of the boundary torque is described with an additional radiative efficiency $\Delta \epsilon$. In the case of NDAF, the gases are cooled via neutrino looses and advection, $\Delta \epsilon$ is no longer a proper parameter matching the extra stresses. Hence, we introduce a factor $\eta$ to quantify the non-zero torque at $r_{\rm{ms}}$:

\begin{equation}
 \textsl{g}_{\rm{ms}}=\eta \dot{M}L_{\rm{ms}},
 \label{eq:1}
\end{equation}
\noindent where $L_{\rm{ms}}=2GM(3\chi_\text{ms}-2a_*)/\sqrt{3}c\chi_\text{ms}$ is the specific angular momentum of a particle in the disk, in which $\chi_\text{ms}=\sqrt{c^2r_\text{ms}/GM}$ (NT73).

In the fluid frame for a time-steady, geometrically thin, relativistic accretion disk, $\eta$ is related to $\Delta \epsilon$ by
\begin{equation}
\eta = \Delta \epsilon \frac{c^2}{\Omega_{\rm ms} L_{\rm{ms}} }.
\label{eq:2}
\end{equation}
For Newtonian disk, this relation is reduced to $\eta=\Delta \epsilon r_{\rm ms}/r_{\rm g}$, where $r_{\rm g} = GM/c^2$ denotes the gravitation radius. The angular velocity of disk at $r_{\rm ms}$ is $\Omega_{\rm ms} = ((r_{\rm ms}^3/GM)^{1/2}+ a_* GM/c^3)^{-1}$.

By using the numerical simulation with Pseudo-Newtonian potential, \citet{HK02} show $\eta\sim\rm{0.05-0.1}$. However, as argued in \citet{K99} and Gammie (1999), in a Kerr metric, the efficiency $\Delta \epsilon$ would become normally greater than unity because the accumulated spin energy of the BH is being tapped. This suggests a link between $\Delta \epsilon$ (as well as $\eta$) and BH spin. As we known, spin energy and angular momentum can be transferred from the BH to the disk via a large-scale closed magnetic field  \citep{B99, van99, LP00, Li00, Li02, Wang02, Wang03, Gan07, Lei07, Lei09}. This mechanism is called magnetic coupling process (MC). We can thus put a constrain on $\eta$ by equating the boundary torque $g_{\rm ms}$ to the total MC torque $T_{\rm MC}$. As shown in Appendix A, $\eta_{\rm max}$ can reach $\sim 10$ for a rapidly spinning BH.

Our nztNDAF model is based on the context given by DPN02, and the general relativistic corrections are adopted from \citet{RH95} (hereafter RH95). The equation for angular momentum for nztNDAF is written as (the details for the derivation are displayed in the Appendix B)

\begin{equation}
\dot{M}r^{2}\sqrt{\frac{GM}{r^{3}}}\frac{D}{A}+\textsl{g}_{\rm{ms}}\frac{A_{ms}}{A}=\textsl{g}=-4\pi r^{2}\tau_{r\varphi}h,
\label{eq:3}
\end{equation}
where $A_{\rm ms}$ is the value of factor $A$ at $r_{\rm ms}$. The second term on the left side of equation (\ref{eq:4}) vanishes for NDAF with the assumption of "zero torque" boundary condition. $A, B, C, D$ and $E$ are the relativistic correction factors for a thin accretion disk around a Kerr BH given by RH95 as,

\begin{equation}
A=1-\frac{2GM}{c^{2}r}+(\frac{GMa_{*}}{c^{2}r})^{2},
\label{eq:4}
\end{equation}

\begin{equation}
B=1-\frac{3GM}{c^{2}r}+2a_{*}(\frac{GM}{c^{2}r})^{3/2},
\label{eq:5}
\end{equation}

\begin{equation}
C=1-4a_{*}(\frac{GM}{c^{2}r})^{3/2}+3(\frac{GMa_{*}}{c^{2}r})^{2},
\label{eq:6}
\end{equation}

\begin{equation}
D=\int_{r_{\rm{ms}}}^{r}\frac{\frac{x^{2}c^{4}}{8G^{2}}-\frac{3xMc^{2}}{4G}+\sqrt{\frac{a_{*}^{2}M^{3}c^{2}x}{G}}-\frac{3a_{*}^{2}M^{2}}{8}} {\frac{\sqrt{rx}}{4}(\frac{x^{2}c^{4}}{G^{2}}-\frac{3xMc^{2}}{G}+2\sqrt{\frac{a_{*}^{2}M^{3}c^{2}x}{G}})}\mathrm{d}x,
\label{eq:7}
\end{equation}

\begin{equation}
E=1-\frac{6GM}{c^2r}+8a_*\left(\frac{GM}{c^2r}\right)^{3/2}-3a_*^2\left(\frac{GM}{c^2r}\right)^2.
\label{eq:8}
\end{equation}

The $\alpha-$ prescription for the viscous shear $\tau_{r\varphi}$, as well as the expression for the disk half-thickness $h$ are corrected as,

\begin{equation}
\tau_{r\varphi}=-\alpha P\frac{A}{\sqrt{BC}},
\label{eq:9}
\end{equation}

\begin{equation}
h=\sqrt{\frac{Pr^{3}}{\rho GM}}\sqrt{\frac{B}{C}},
\label{eq:10}
\end{equation}
where $P$ is the total pressure, including gas pressure $P_{\rm gas}$, radiation pressure $P_{\rm rad}$, degeneracy pressure $P_{\rm deg}$, neutrino pressure $P_{\nu}$ and the magnetic pressure $P_{\rm B}$:

\begin{equation}
P=P_{\rm{gas}}+P_{\rm{rad}}+P_{\rm{deg}}+P_{\rm{\nu}}+P_\text{B},
\label{eq:11}
\end{equation}

\noindent here, we assume that the magnetic pressure accounts for a fraction of the total pressure as $P_{\rm B}=\beta P$. Other terms are expressed in Appendix C.

According to \cite{RH95} (see their equation 19), the viscous heating rate is

\begin{equation}
Q_{\rm vis}^+=\frac{3}{4}\sqrt{\frac{GM}{r^3}}\frac{A}{B}\int_{-h}^{h}\tau_{r\phi}\mathrm{d}z,
\label{eq:12}
\end{equation}

\noindent substituting equation (\ref{eq:B13}) into equation (\ref{eq:12}), we have

\begin{equation}
Q_{\rm{vis}}^+=\frac{3GM\dot{M}}{8\pi r^{3}}\frac{D}{B}+\frac{3g_\text{ms}}{8\pi r^2}\sqrt{\frac{GM}{r^3}}\frac{A_\text{ms}}{B},
\label{eq:13}
\end{equation}
where the factor $D/B$ is equal to zero at $r_{\rm{ms}}$ and approaches unity at large radii. As stated in \citet*{J10}, this asymptotic behaviour of $D/B$ is the same as for the boundary condition derived in NT73 and \citet{CB07}, who used more complex formalism in the Kerr metric. The second term on the right side of the equation is the contribution of the none-zero torque at $r_\text{ms}$. This term is non-zero at the inner edge, which will increase the disk luminosity.

The equation for the energy balance is

\begin{equation}
Q_{\rm{vis}}^+ =Q_{\rm{\nu}}^- +Q_{\rm{photo}}^- +Q_{\rm{adv}}^-
\label{eq:14}
\end{equation}
where $Q_{\rm{\nu}}^-$ is the total cooling rate due to neutrino losses, $Q_{\rm{photo}}^-$ is the photodisintegration and $Q_{\rm{adv}}^-$ the advective cooling rate. Detailed expressions for $Q_{\rm{photo}}^-$, $Q_{\rm{adv}}^-$ and the bridging formula for $Q_{\rm{\nu}}^-$ are given in DPN02 (see also Appendix C).

\section{The properties of non-zero torque NDAF model}
\label{sect:results}
We are interested primarily in the properties of the inner accretion flow, where the neutrino process is important. As argued in PWF99, NPK01 and DPN02, the flows are fully advection-dominated for $r>100r_{\rm{g}}$, where neutrino cooling is not important and photons are completely trapped. Therefore, we focused on the region from $r_{\rm{ms}}$ to $r_{\rm{max}}=100r_{\rm{g}}$. In the calculation, we do not include the cooling term arising from the photodisintegration $Q_{\rm{photo}}^-$ because it is much less than the neutrino cooling rate in the inner region (\citealt{J04}). The strength of the non-zero torque is described by the parameter $\eta$ referred in equation (\ref{eq:1}). Throughout the paper, we take $\alpha=0.1$ as a typical value, for the detailed effects of $\alpha$ one can refer to previous studies (e.g., \citealt{CB07, Liu10b, Lin16}).

\subsection{The structure of nztNDAF}
\begin{figure*}[h!tp]
\centering
\includegraphics[width=\textwidth]{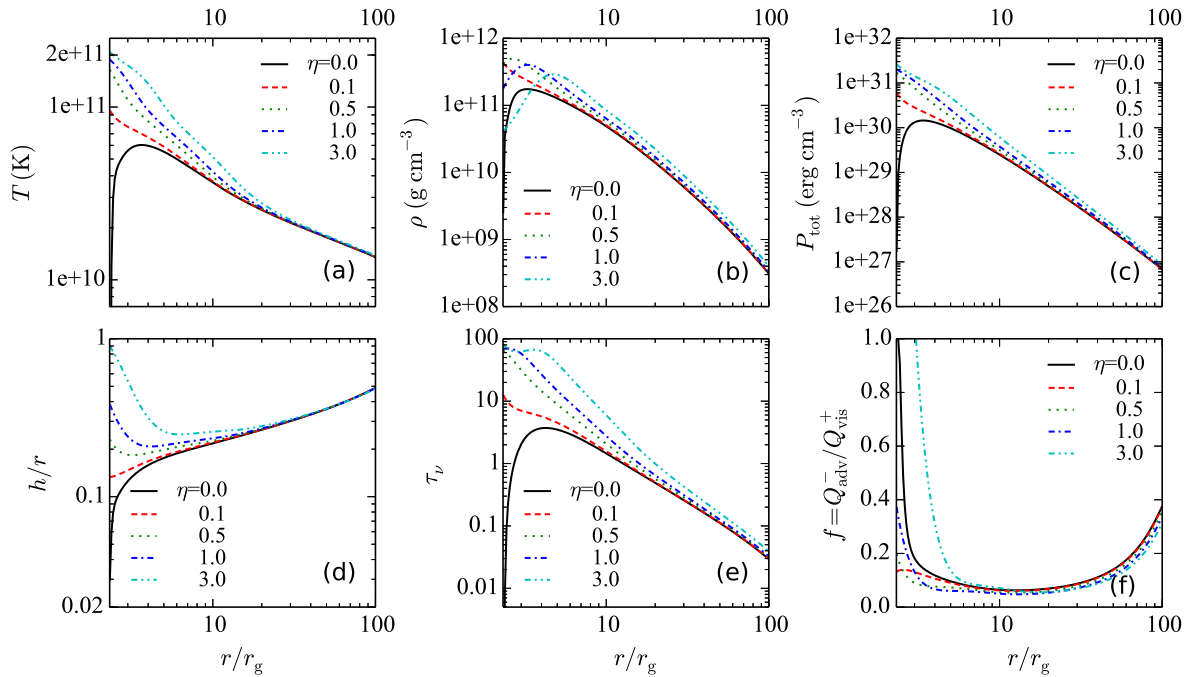}
\caption{NDAF solutions for a viscosity parameter $\alpha=0.1$, accretion rate $\dot{m}=1.0$ ($\dot{m} \equiv \dot{M}/\rm{M_{\sun}\,s^{-1}}$), BH mass $M=7\, \rm{M_{\odot}}$ and spin $a_{*}=0.9$. The seven panels show (a) the temperature $T$, (b) density $\rho$, (c) total pressure $P$, (d) disk height $H$, (e) total neutrino optical depth $\tau_\nu$ and (f) advection parameter ($f=Q_{\rm{adv}}/Q_{\rm{vis}}$) as a function of the disk radius, for four $\eta$ values: $\eta=0.0$(black solid lines), 0.1 (red dashed lines), 0.5 (green dotted lines), 1.0 (blue dashdotted lines), 3.0 (cyan dashdotted lines). Curves for $\eta=0$ correspond to the solutions for previous NDAF model with zero-torque boundary.}
\label{struct}
\end{figure*}

We solve numerically equations (\ref{eq:3}) -- (\ref{eq:16}) to find the disk temperature $T$ and density $\rho$ versus the disk radius with a typical model parameters $\alpha=0.1$, $M=7\, \rm{M_{\odot}}$, $a_{*}=0.9$, $\dot{m}=1.0$ ($\dot{m} \equiv \dot{M}/\rm{M_{\sun}\,s^{-1}}$). The solutions are shown in Figure \ref{struct}. In order to study the effects of the boundary torque, we calculate the solutions for $\eta=0$ (black solid lines), 0.1 (red dashed lines), 0.5 (green dotted lines), 1.0 (blue dashdotted lines), 3.0 (cyan long dashed lines). Curves with solids lines ($\eta=0$) in Figure \ref{struct} exhibit the solutions for NDAF without boundary torque. In Appendix C, we have made an effort to obtain the analytic solutions of nztNDAF for better understanding the main results the numerical calculation exhibited here.

From Figure \ref{struct}, we find that the boundary torque has strong effects on the properties of inner disk. For example, as shown in figures 1a, 1b, 1c ,1d and 1e, the temperature $T$, density $\rho$, pressure $P$, height $h$ and neutrino optical depth $\tau_\nu$ become non-zero at $r_{\rm ms}$ due to the existence of such boundary torque. According to equation (15), a disk with a greater boundary torque will produce more heat in the inner region, leading to a higher temperature as shown in figure 1a. As a result, the disk pressure $P$, height $h$ and neutrino optical depth $\tau_\nu$ increase with the increasing $\eta$. In figure 1f, the drop of advection parameter $f=Q_{\rm adv}/Q_{\rm vis}$ in inner region reflects that this additional heating indeed ignites efficient neutrino cooling. However, as discussed in DPN02, the cooling rate due to neutrino emission will be suppressed if $\tau_{\rm{\nu}}$ is too large. This is also illustrated in figure 1f. The advection becomes important ($f>0.5$) in inner region if $\eta$ becomes significantly larger than 1. From figure 1, we also find that the boundary torque weakly affect the outer disk. This is because the additional heating term (the second term in the right side of equation (15)) scales as $r^{-7/2}$ at large $r$ rather than $r^{-3}$ as in the standard viscous heating term (the first term).

\begin{figure*}[h!tp]
\centering
\includegraphics[width=60mm]{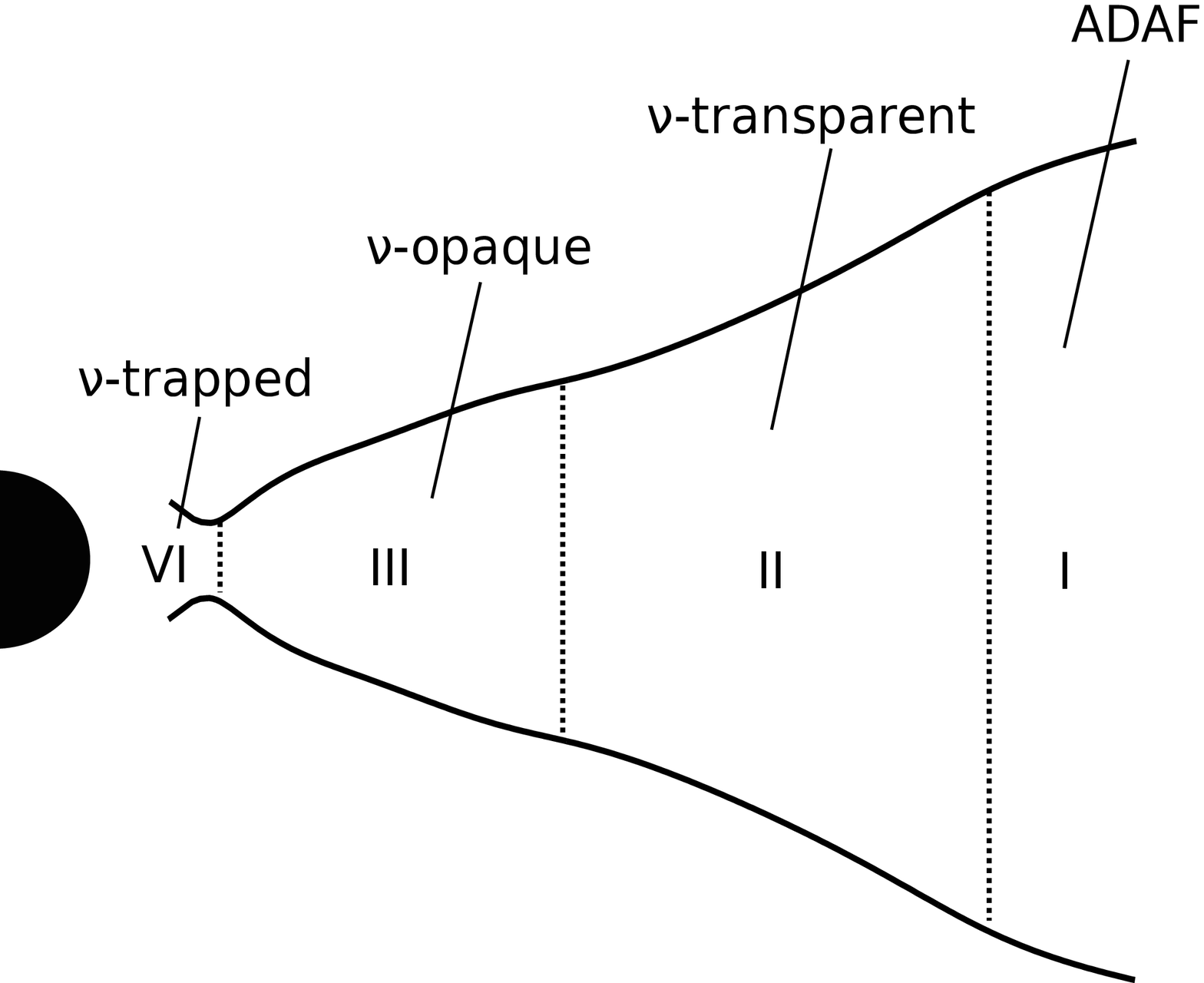}
\includegraphics[width=60mm]{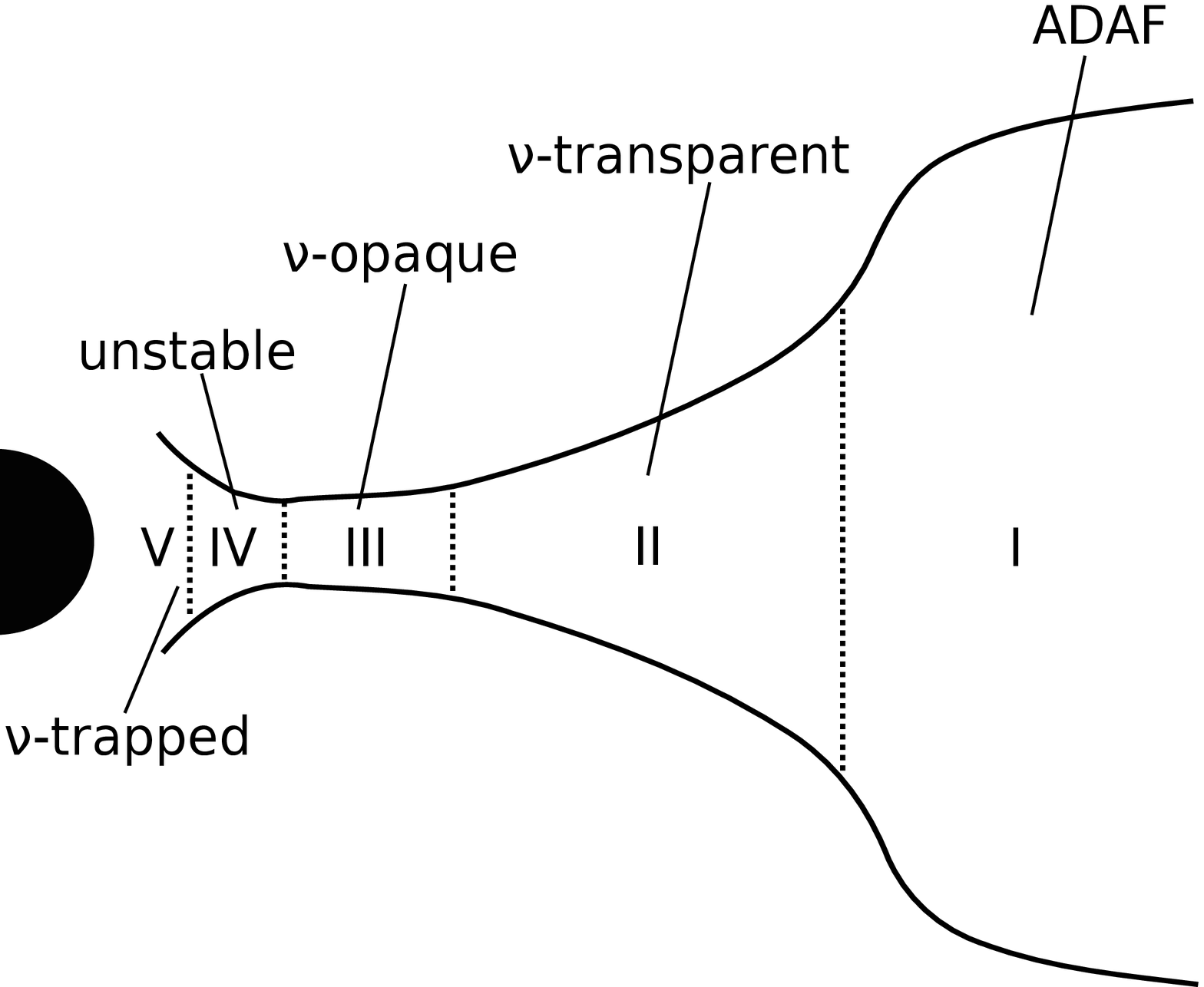}
\caption{Schematic picture of NDAF (left) and nztNDAF (right). Detailed explanations and analytical solutions for these regions are given in Appendix C.}
\label{sketch}
\end{figure*}

It is shown in Figure \ref{struct}f that advection (denoted by $f$) dominates at large radii for both NDAF and nztNDAF. An equivalent statement is that the cooling timescale is much longer than the accretion timescales, so the energy is advected inward before it can be radiated away. As shown in figure 1f, the advection parameter $f$ slowly decreases as the gas continues to fall inward, since the increasing temperature and density produce a rapid increase in the neutrino cooling rate. For NDAF, the densities and temperatures near inner edge are too small to ignite significant neutrino cooling, and then $f$ goes to unity again. So as discussed in Appendix C, NDAF generally consists of four regions as shown in Figure \ref{sketch} (see also Figure 10 in \citet{CB07}:

(I) at large radii, densities and temperatures are too small for neutrino cooling to be significant, and the disk is simply an advection-dominated flow (ADAF).

(II) at this region, neutrino emission switches on. The neutrino opacity is not important. So this region is referred as transparent NDAF.

(III) disk becomes opaque for neutrinos, but neutrino cooling is still dominated. We call this region as opaque NDAF.

(VI) near $r_{\rm ms}$, the flow returns to ADAF due to the low temperature and density.

For nztNDAF, the inner structure are quite different with NDAF. There will be two regions inside region III:

(IV) in this region, the temperature is very high due to the additional heating driven by boundary torque. The disk is thus dominated by radiation pressure,  but still a opaque NDAF. In section 3.2, we will show that this region is viscously unstable. For this reason, it is named unstable NDAF.

(V) since huge heat is produce near $r_{\rm ms}$, the neutrino optical depth is so high that even neutrinos can not escape any more, resulting in an advection cooling flow (corresponding to $r<4r_{\rm g}$ for $\eta =3.0$ in Figure 1f).

The analytical solutions for each region are given in detail in Appendix C. To show the goodness of these analytical solutions, we compare them with the numerical ones. In Figures \ref{Fig:C1} and \ref{Fig:C2} of Appendix C, it is clearly shown that they are consistent with our numerical solutions. These studies suggest that the analytical solutions can capture the main feature of the disk.

\subsection{Stability Analysis: Viscous Instability}

NDAF are said to be stable under most cases (NPK01, DPN02). As shown in the section 3.1, by introducing the boundary torque, the nztNDAF behaviours quite different with NDAF. The inner disk will become viscously unstable if a strong magnetic stress applied on its edge. One can refer to Appendix C for a better understanding of this statement. For disk with high accretion and large $\eta$, the temperature is significantly increased due to the additional heating driven by boundary torque. As a result, the flow becomes radiation pressure and neutrino pressure dominated, which is unstable according to the viscous instability criterion $d\dot{m}/d\Sigma <0$ ($\Sigma$ is the surface density). As an example, for $\dot m=1.0$ and $\beta=0$, the instability will occur when $\eta\gtrsim0.45$ (refer to Figure \ref{Fig4}). The  corresponding magnetic field near $r_{\rm ms}$ will be $\gtrsim 4.6 \times 10^{15}\ \ \mathrm{Gauss}$, which is estimated by equating the magnetic torque $\eta \dot M L_{\mathrm{ms}}$ to $2\pi r_{\mathrm ms}^2 \cdot 2h_{\mathrm ms}\cdot\frac{\langle B\rangle^2}{4\pi}$, where $\langle B\rangle$ denotes the magnitude of the magnetic field.

\begin{figure}[h!tp]
   \centering
   \includegraphics[width=60mm]{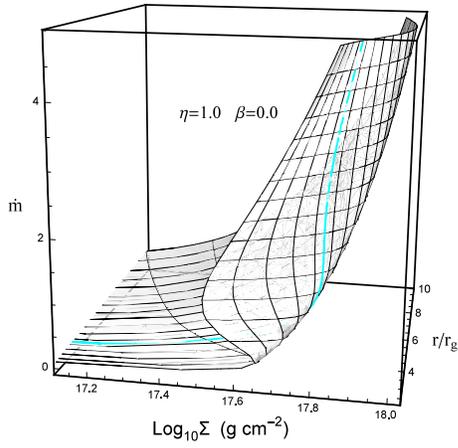}\\
   \caption{The $\dot{m}-\Sigma$ profile at different disk radius $r$. The thick cyan curve denotes the last viscously stable radius $r_{\rm ur}$ for any $\dot{m}$, which is located at $3.43 r_\mathrm{g}$.}
   \label{Fig3}
   \end{figure}

\begin{figure*}[h!tp]
   \centering
   \includegraphics[height=60mm]{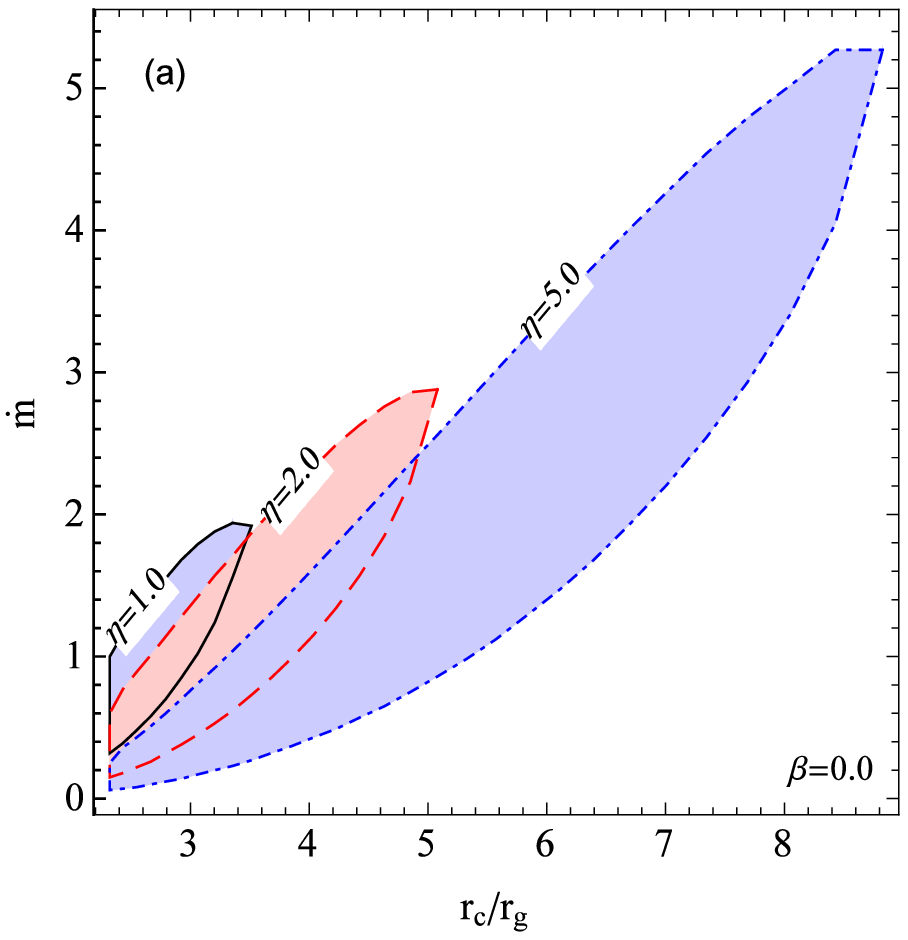}
   \includegraphics[height=60mm]{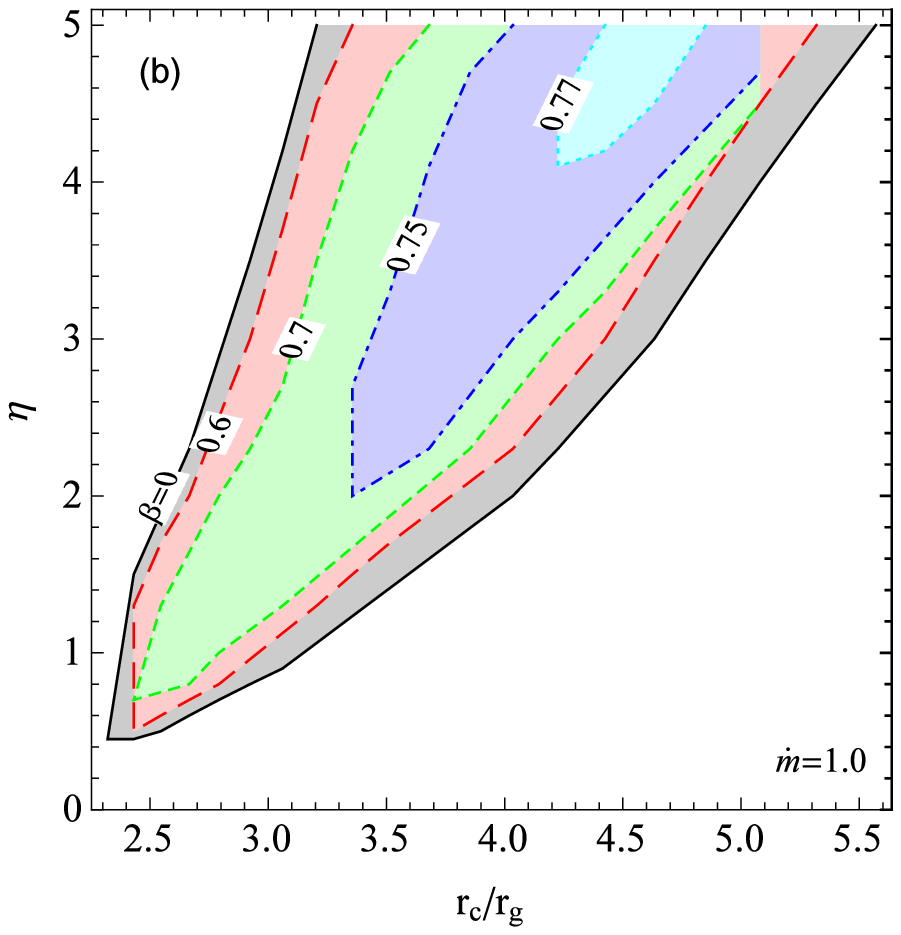}
   \caption{(a). The viscous unstable regions are indicated by the shaded regions for different inner edge torque $\eta=1.0$, $2.0$, and $5.0$, with $M=7\, \rm{M_{\odot}}$, $a_{*}=0.9$ and $\dot{m}=1$. Note that for each fixed disk radius in the unstable region, there are two critical values of accretion rate $\dot{m}_{cr,l}$ and $\dot{m}_{cr,u}$, the flow at $r$ will be unstable when $\dot{m}_{cr,l}<\dot{m}<\dot{m}_{cr,u}$. The upper and lower border lines of each of the shaded region separately denote the critical accretion rate $\dot{m}_{cr,u}$ and $\dot{m}_{cr,l}$ for different disk radius. (b). The right panel plots the unstable region versus $\eta$ for different magnetic pressure component $\beta=0.0$, $0.6$, $0.7$, $0.75$,and $0.77$, with fixed $\dot m =1.0$. Note that the unstable region is only significantly affected by $\beta$ when it is is greater enough, like $\beta > 0.6$.}
   \label{Fig4}
   \end{figure*}

Figure \ref{Fig3} shows the $\dot{m}-\Sigma$ profile for different radius $r$ and different $\eta$. The cyan line is a critical radius $r_{\rm ur}$ beyond which the solution will be stable for all accretion rate $\dot{m}$. For $\dot{m}=1.0$ and $\beta=0.0$, this cyan line locates at $r=3.43r_{\rm g}$, but it may vary with $\eta$ and $\beta$.

Figure \ref{Fig4} shows the unstable region which depends on $\dot m$, $\eta$ and $\beta$, and the unstable zones are shown as the shaded regions. For fixed $\eta$ and $\dot{m}$, the disk might only be unstable in a radius range. With increasing $\eta$ and $\dot{m}$, the unstable region moves further out in the flow. Take the scenario $\dot{m}=1.0$ and $\beta=0.0$ as an example, the whole disk is viscously stable if $\eta$ is not too large (not greater than 0.45), however, the disk will become viscously unstable in the inner region $r<2.5r_{\rm g}$ when  $\eta \sim 0.5$. In addition, the unstable region expands outwards to $\sim3r_{\rm g}$ if $\eta\sim1$. Note that there is a critical radius $r_{\rm ur}$ for the unstable region for each $\eta$, for example, the cyan line in figure \ref{Fig3}. At $r<r_{\rm ur}$, the viscous instability take place only in a certain accretion rate range, like $0.32<\dot{m}<1.94$ if $\eta=1.0$. Therefore, one conclusion is that under a larger inner edge torque, the disk can be viscously stable only if the accretion rate is relatively low or extremely high, while the disk with a moderate accretion rate may suffer instability. This statement can be understood as follow: if the accretion rate is relatively low, the temperature can not be high enough and consequently the radiation pressure can not take the dominant role; on the other hand, if the accretion is extremely high, then the neutrinos will be trapped due to the extremely high neutrino optical depth and the flow will become advection cooling dominated, both of those two scenarios can't accord with the unstable condition.

\subsection{Neutrino Annihilation Luminosity}

Inspecting equation (15), the non-zero torque applied on the inner edge results in huge energy dissipation, which would lead to a more powerful neutrino radiation as well as a greater neutrino annihilation luminosity. The total neutrino luminosity from the accretion flow is expressed as
\begin{equation}
L_{\rm{\nu}}=4\pi\int_{r_{\rm{ms}}}^{r_{\rm{max}}}Q_{\rm{\nu}}^-r\mathrm{d}r
\label{eq:15}
\end{equation}
where we adopt $r_{\rm{max}}=100r_{\rm{g}}$ as discussed above. Our method for calculating neutrino annihilation is similar to PWF99 and \citet{R03}. The disk is modelled as a grid of cells in the equatorial plane. A cell $k$ has its neutrino mean energy $\varepsilon_{\nu_{i}}^{k}$ and luminosity $l_{\rm{\nu}_{i}}^{k}$, and the height above (or below) the disk is $d_{k}$. The angle at which neutrinos from cell $k$ encounter antineutrinos from another cell $k'$ at that point is denoted as $\theta_{kk'}$. Then the neutrino annihilation luminosity at that point is given by the summation over all pairs of cells,

\begin{eqnarray}
l_{\rm{\nu\bar{\nu}}}=A_{1}\sum_{k}\frac{l_{\rm{\nu}_{i}}^{k}}{d_{k}^{2}}\sum_{k'}\frac{l_{\rm{\nu}_{i}}^{k}}{d_{k}^{2}} (\varepsilon_{\nu_{i}}^{k}+\varepsilon_{\bar{\nu_{i}}}^{k'})(1-\cos\theta_{kk'})^{2} \nonumber \\ +A_{2}\sum_{k}\frac{l_{\rm{\nu}_{i}}^{k}}{d_{k}^{2}}\sum_{k'}\frac{l_{\rm{\nu}_{i}}^{k}}{d_{k}^{2}} \frac{\varepsilon_{\nu_{i}}^{k}+\varepsilon_{\bar{\nu_{i}}}^{k'}}{\varepsilon_{\nu_{i}}^{k}\varepsilon_{\bar{\nu_{i}}}^{k'}} (1-\cos\theta_{kk'})
\label{eq:16}
\end{eqnarray}
where $A_{1}\approx1.7\times10^{-44}\rm{cm\, erg^{-2}\, s^{-1}}$ and $A_{2}\approx1.6\times10^{-56}\rm{cm\, erg^{-2}\, s^{-1}}$.

The total neutrino annihilation luminosity is obtained by integrating over the whole space outside the BH and the disk,

\begin{equation}
L_{\rm{\nu\bar{\nu}}}=4\pi\iint l_{\rm{\nu\bar{\nu}}}r\mathrm{d}r\mathrm{d}z
\label{eq:17}
\end{equation}

\begin{figure}[!ht]
   \centering
   \includegraphics[height=60mm]{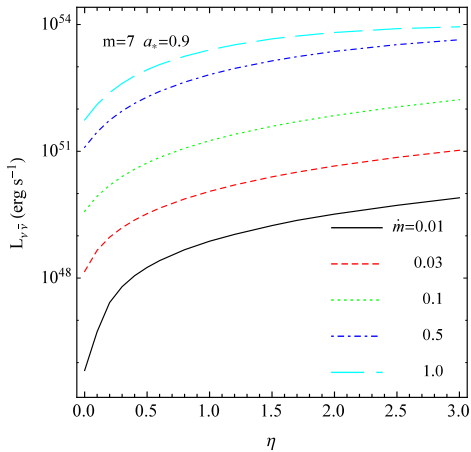}
   \caption{ $L_{\rm{\nu\nu}}$ vs. $\dot{m}$. Other parameters are $M=7{M_{\odot}},\ a_*=0.9,\ \alpha=0.1,\  \beta=0$.}
   \label{Fig5}
   \end{figure}

\begin{figure*}[!ht]
   \centering
   \includegraphics[height=100mm]{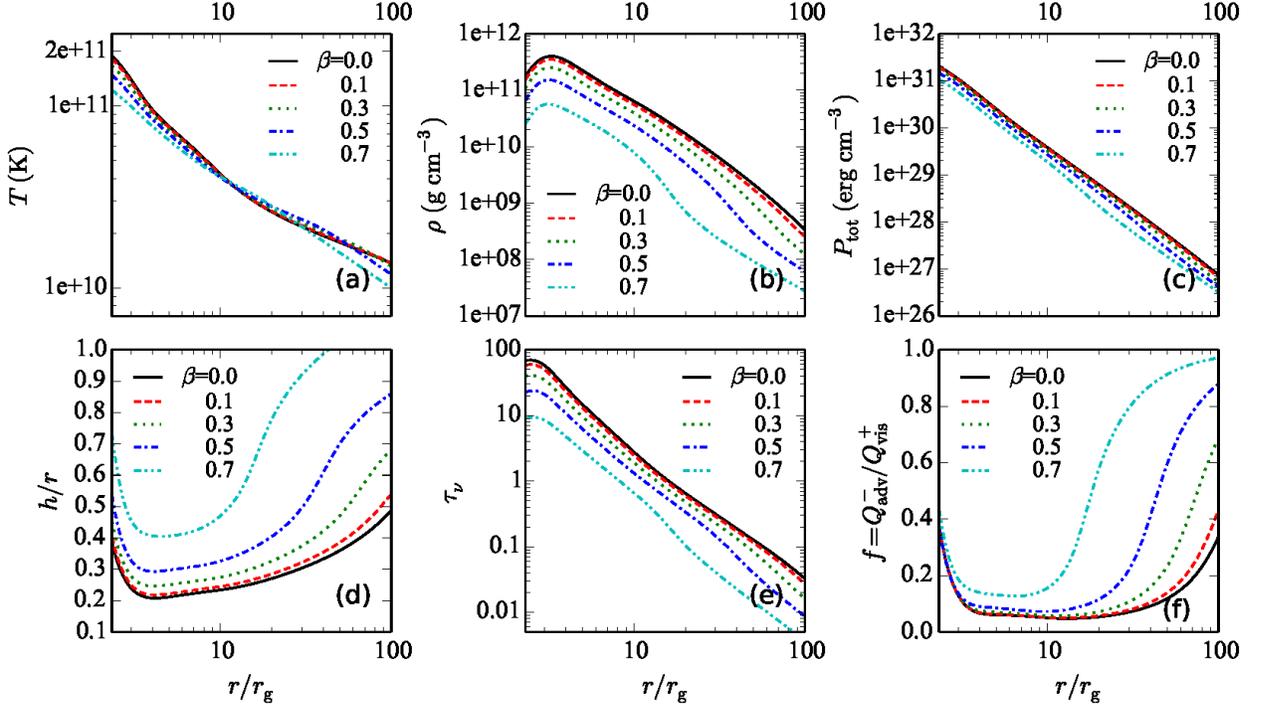}
   \caption{Same to Figure \ref{struct}, but focused on the effect of the magnetization parameter $\beta$. The six panels separately show (a) the temperature $T$, (b) density $\rho$, (c) total pressure $P$, (d) the ratio of disk height to radius $h/r$, (e) total neutrino optical depth $\tau_\nu$ and (f) advection parameter ($f=Q_{\rm{adv}}/Q_{\rm{vis}}$) as a function of the disk radius, for five $\beta$ values: $\beta=0.0$(black solid lines), 0.1 (red dashed lines), 0.3 (green dotted lines), 0.5 (blue dashdotted lines), 0.7 (cyan dashdotted lines). The rest parameters are fixed, i.e., the inner edge torque $\eta=1.0$, the viscosity $\alpha=0.1$, the accretion rate $\dot{m}=1.0$ ($\dot{m} \equiv \dot{M}/\rm{M_{\sun}\,s^{-1}}$), the BH mass $M=7\, \rm{M_{\odot}}$ and spin $a_{*}=0.9$.}
   \label{Fig6}
   \end{figure*}

Figure \ref{Fig5} demonstrates the variation of $L_{\rm{\nu\bar{\nu}}}$ versus $\dot{m}$ with $\eta=$0 (dashed line), 0.5 (dotted line), 1,3(solid line). It is found that the neutrino annihilation luminosity is greatly strengthened by increasing the torque on the inner edge. According to our calculation, $L_{\rm{\nu\bar{\nu}}}$ varies from $1.9\times10^{51}\rm{erg\, s^{-1}}$ to $3.8\times10^{54}\rm{erg\, s^{-1}}$ for $\eta=1.0$, we find that the $L_{\rm{\nu\bar{\nu}}}$ stay constant around $10^{54}\rm{erg\, s^{-1}}$ for accretion rate  $\dot{m}\approx5.0$. This implies that the effect of neutrino optical depth becomes important. Figure \ref{Fig6}(b) shows the variation of $L_{\rm{\nu\bar{\nu}}}$ versus $\eta$ with different mass accretion rates $\dot{m}(0.01, 0.03, 0.1, 0.5, 1.0)$.

\begin{figure}[!ht]
\centering
\includegraphics[width=70mm]{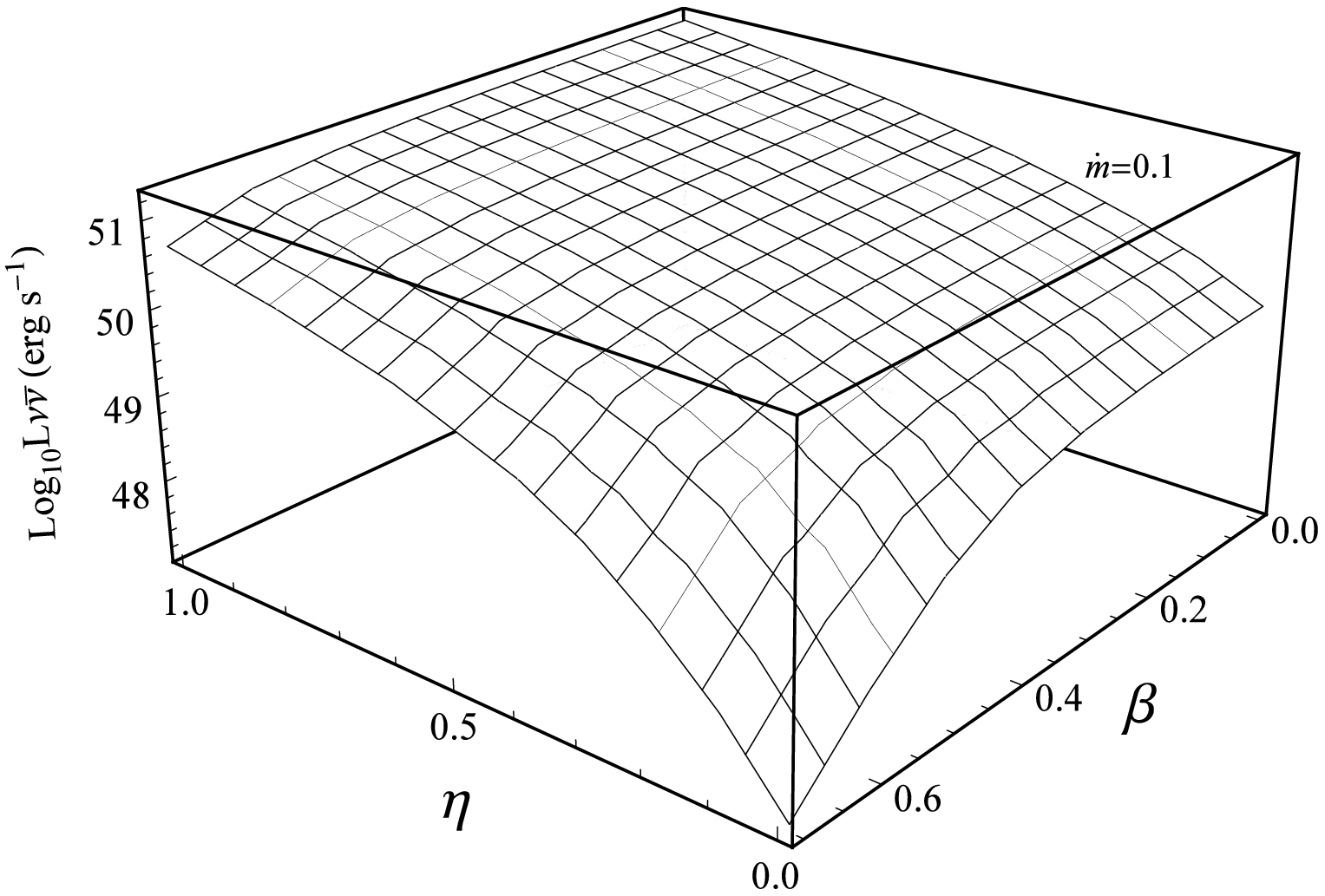}
\caption{The neutrino annihilation luminosity $L_{\rm{\nu\bar{\nu}}}$ versus the inner edge torque parameter $\eta$ and accretion flow's magnetic pressure parameter $\beta$ for $\dot{m}=0.1$.}
\label{Fig7}
\end{figure}

We have fixed the magnetic pressure parameter $\beta$ to zero so far, this parameter could be important and deserves discussion. A consideration is that the accretion flow might be magnetized, and the magnetic pressure can accounts for certain part of the total pressure. Figure \ref{Fig6} shows the structure of the disk with different magnetization $\beta$. We find that the disk will become thicker when a significant magnetic pressure($\beta\gtrsim0.5$) is involved. Consequently, the viscous instability and neutrino luminosity are expected to be suppressed by the strong magnetic pressure, as illustrated by Figure \ref{Fig4}(b) and Figure \ref{Fig7}. However, if $\beta$ is not too large ($\beta\lesssim0.3$), the structure and the neutrino annihilation luminosity of the disk are weakly affected. For $\eta <10$, our estimated magnetic parameter $\beta$ is generally less than 0.3 (see also the Figure 6 in \cite{Cao14}). Therefore, we just take $\beta=0$ as a good approximation.

\section{Interpreting the luminosities of Bright SGRBs, LGRBs and ULGRBs}
%\section{The inner edge torque required by some short GRBs}
The prevailing opinion about the progenitors of short GRBs (SGRBs) and long GRBs (LGRBs) is that they are separately the results of compact binaries merger and the collapse of massive stars. For accreting BH central engine, the limited total material mass that can be supplied during such two types of events sets a concrete constraint to the accretion model. The related work from this perspective has been carried out by previous authors \citep{Liu15, Song16}. Since the relative low output power and corresponding unreasonable high requirements of the total amount of accreting mass, the NDAF model is challenged to interpret certain bright SGRBs and powerful LGRBs. In this section, we will show that once the the inner edger torque is considered, the NDAF model still works well for those ``problem samples" pointed out by previous authors. Meanwhile, using the same method of \cite{Liu15}, we investigate our nztNDAF model for ultra-long GRBs (ULGRBs) in the frame of BSG-progenitors.

\emph{\textbf{---SGRBs}}: Based on the limitation for the accretion disk mass after the compact binaries coalescence given by numerical simulations \citep{RJ98, RJ01, KL98, LK99, P99, Liu12}, \cite{Liu15} argued that some SGRBs may could not be explained by the common NDAF model, since the mass of remanent disk required by the model significantly exceeds the reasonable range which is given by previous numerical simulations. Specifically speaking, for neutron star binaries merging (NS+NS), the reasonable mass of the remaining disk is likely in the range of $0.1-0.2 M_\odot$ \citep{RJ98, RJ01}, while for the coalescence of neutron star and black hole (NS+BH), the survived disk mass is not likely larger than $0.5 M_\odot$ \citep{KL98, LK99, P99, Liu12}. However, as shown in \cite{Liu15}, for several SGRBs (such as GRB050724, 051221A, 090426 and 120804A), if a common NDAF disk is taken as the model of central engine, then the disk mass constrained by combining the observational data with the common NDAF model will easily exceed the limiting value above for a large range of parameters such as $m$ and $a_*$, except for some extreme values of those parameters, i.e. an extremely low BH mass $m$ or an extremely high spin $a_*$.

As discussed in section 3.2, the neutrino annihilation luminosity of nztNDAF is much larger than NDAF without inner edge torque, the energy problem addressed above may be solved by nztNDAF model. The same observed luminosity will requires a smaller mass accretion rate and consequently a smaller disk mass under a fixed time duration in nztNDAF. Here, we will estimate the value of $\eta$ required to fit those ``problem SGRBs''.

Considering the conversion efficiency, the output power from the NDAF central engine is calculated as follow:
\begin{equation}
\dot{E}=\eta_{\nu \bar{\nu}} L_{\nu \bar{\nu}} \label{eq:18}
\end{equation}
in which $\eta_{\nu \bar{\nu}}$ is the conversion factor \citep{AJM05, FW11, Liu12, Liu15}. Meanwhile, from the point view of observation, the output power can be evaluated as follow:
\begin{equation}
\dot{E}\approx \frac{(1+z)(E_{\gamma,\mathrm{iso}}+E_{\mathrm{k,iso}})\theta_\mathrm{j}^2}{2T_{90}} \label{eq:19}
\end{equation}
where $E_{\gamma,\mathrm{iso}}$ is the isotropic energy of prompt emission, $E_{\mathrm{k,iso}}$ is the isotropic kinetic energy of the fireball constrained by fitting the afterglow emission, $z$ is the redshift, $T_{90}$ is the time duration of the prompt emission, $\theta_\mathrm{j}$ is the jet angle.

For a typical set of parameters ($m$, $a_*$, $\eta$, $\eta_{\nu \bar{\nu}}$), we can estimate the mass accretion rate and then the disk mass $m_\mathrm{disk}=\dot{m} T_{90}/(1+z)$ by adopting eq.(\ref{eq:18}) to a GRB ($E_{\gamma,\mathrm{iso}}$, $E_{\mathrm{k,iso}}$, $\theta_\mathrm{j}$, $T_{90}$, $z$).

For comparison, the required values of mass of disk based on the nztNDAF model ($\eta\neq0$) and NDAF ($\eta=0$) are listed in Table 1 for four ``problem SGRBs'' referred above. We study two possible scenarios, i.e., NS+NS merging and NS+BH merging, and adopt different BH mass and maximum disk mass. For NS+NS merging, $m_{\rm disk}<0.2 M_{\sun}$ and $m=3$. For NS+BH merging, we take larger values, i.e., $m_{\rm disk}<0.5 M_{\sun}$ and $m=7$. Meanwhile we take a moderate BH spin  $a_*=0.5$. The results are summarized in the top panel of Table 1.

From Table 1, we conclude that the inner boundary torque can exactly reduce the requirement of mass of disk so as to make it come back to the reasonable range, and solve the problem addressed in \cite{Liu15} (which uses a  traditional NDAF model). Taking GRB 050724 as an example, for NS+BH case with NDAF model , the required disk mass should be $1.54 M_\odot$, which is much larger than the upper limit $0.5 M_\odot$ (see \cite{Liu15} for the same conclusion). However, if we take nztNDAF model with $\eta=0.23$, $m_{\rm disk}$ can be reduced to the reasonable value $0.5 M_\odot$. The disk mass can be even smaller if we increase $\eta$. The analyses to other samples are similar, and nztNDAF works well for those SGRBs.

\emph{\textbf{---LGRBs}}: Using the similar method of \cite{Liu15} for SGRBs, \cite{Song16} investigated the mass distribution of the NDAF disk for 48 LGRBs, $5 M_\odot$ is thought to be an fiducial amount of the remanent material of massive collapsars. Their work showed that NDAF may not be suitable for some extremely high energy LGRBs because they require a unusually large disk mass $>5 M_\odot$. Here, we fit these ``problem LGRBs'' with nztNDAF, the results are listed in the middle panel of Table 1. BH mass $m=3$ and spin $a_*=0.9$ are adopted in the fits.

As shown in Table 1, for most LGRBs, the values of $\eta$ are smaller than unity, except for GRB 050820A. However, $\eta=1.14$ is only slightly larger than 1. We thus conclude that all these energetic LGRB can be well fitted by nztNDAF .

\emph{\textbf{---ULGRBs}}: The blue supergiants (BSGs) are considered as possible progenitors of ULGRBs which possess time duration of about $10^4$ seconds or even longer \citep{NKSN13}. The masses of the BSGs are about several tens of to hundreds of solar mass. Here we investigate the possibility of powering ULGRBs with NDAF. Due to low metallicity, the progenitor envelope are considered to form BH without significant mass loss \citep{Heger13}, hereby we take $50 M_\odot$ as the reference value of the accreted material. The core may be rapidly rotating at collapse, we thus set BH spin as $a_* =0.9$. For simplicity, we take BH mass $3 M_\odot$ without considering its gradually growing up. \cite{NKSN13} also assumed that the relativistic jet is lunched after the mass of the BH reaching to $3 M_\odot$. With this BH mass, our fits may only draw a lower limit on disk mass, since the neutrino annihilation luminosity $L_{\nu\bar{\nu}}$ is anti-proportional to the BH mass.

The fitting results for 3 ULGRBs are listed in Table 1. We find that NDAF is not suitable for ULGRBs because the needed disk mass seriously exceed $50 M_\odot$. One thus needs our nztNDAF model to interpret ULGRBs. Note that the fit for GRB 111209A require a rather large boundary torque $\eta \simeq 3$. As shown in section 2 and Appendix A, the maximum value of boundary torque parameter $\eta_{\rm max}$ can be greater than 10. So, this large $\eta$ is still acceptable.

\begin{table*}
\label{Tab:table1}
\centering
\begin{minipage}{\textwidth}
\caption{The inner edge torque parameter $\eta$ constrained by accretion disk mass for different type of GRBs}
\begin{tabular*}{\textwidth}{@{\extracolsep{\fill}}llllllllll}
  \hline\noalign{\smallskip}\hline\noalign{\smallskip}
  GRB & Reference & $z$ & Duration & $E_{\gamma,\mathrm{iso}}$ & $E_{\mathrm{k,iso}}$ & $\theta_\mathrm{j}$ & m & $\eta$ & $m_{\rm{disk}}$ \\
      &   &   &   (s)    & ($10^{51}\ \mathrm{ergs}$)&($10^{51}\ \mathrm{ergs}$)&        (rad) &($M_\odot$)&   & ($M_\odot$)      \\
  \hline\noalign{\smallskip}
  050724  & 4 & 0.257 & 3   & 0.1   & 0.27  & $\gtrsim$0.35 & 7 & 0.23 & 0.5 (1.54) \\
          & 4 &       &     &       &       &               & 3 &  0.36 &  0.2 (0.88) \\
  051221A & 4 & 0.5465& 1.4 & 0.92  & 12.6  & $\sim$0.12    & 7 & 0.28 & 0.5 (1.72) \\
          & 4 &       &     &       &       &               & 3 &0.44 & 0.2 (0.98) \\
  090426  & 4 & 2.609 & 1.2 & 2.84  & 135   & $\sim$0.07    & 7 & 0.23 &  0.5 (1.72) \\
          & 4 &       &     &       &       &               & 3 & 0.44 & 0.2 (0.97) \\
  120804A & 4 & 1.3   & 0.81& 3.88  & 56.9  & $\gtrsim$0.19 & 7 & 0.59 & 0.5 (2.99) \\
          & 4 &       &     &       &       &               & 3 & 0.99 &  0.2 (1.7) \\
  \noalign{\smallskip}\hline
  990123  & 6 & 1.600 &63.30$\pm$0.26&1437.9$\pm$177.8 &202.8$\pm$18.5 & 0.086$\pm$0.0075   &3&  0.17 & 5 (10.6) \\
  021004  & 6 & 2.3304&77.1$\pm$2.6  & 55.6$\pm$7.2    & 83.5$\pm$14.5 & 0.221$\pm$0.0787    &3 & 0.07 & 5 (7.97) \\
  050820A & 6 & 2.6147 & 128.0$\pm$106.9 & $970^{+310}_{-140}$ & $5370^{+800}_{-950}$ &$0.1152^{+0.0087}_{-0.0052}$ & 3 & 1.14 & 5 (30.9) \\
  060124  & 6 & 2.297 & 298$\pm$2 & 420$\pm$50 & $5788.7^{+1107.9}_{-126.6}$ & $0.0530^{+0.0091}_{-0.0040}$ & 3 & 0.93 & 5 (25.8) \\
  060210  & 6 & 3.9133 & 220$\pm$70 & 353$\pm$19 & $11132.9^{+1053.9}_{-947.2}$ & $0.0209^{+0.0030}_{-0.0021}$ & 3 & 0.17 &  5 (10) \\
  %  050904  & 9 & 6.295  & 183.6$\pm$13.2  & $1325.5^{+678.2}_{-400.7}$ & $884^{+863}_{-442}$ & 0.034$\pm$0.0051 &3& 0, 2.7e-3 & 5.4, 5 \\
  070125  & 6 & 1.5477 & 63.0$\pm$1.7 & $957.6^{+106.4}_{-87.4}$ & $64.3^{9}_{1.7}$ & 0.23$\pm$0.0105 & 3 & 0.67 & 5 (20.8) \\
  090323  & 6 & 3.568 & 133.1$\pm$1.4 & 3300$\pm$130 & $1160^{+130}_{-90}$ & $0.0489^{+0.0069}_{-0.0017}$ & 3 & 0.20 &  5 (11.1) \\
  090926A & 6 & 2.1062 & 20$\pm$2 & $1890\pm30$ & $68\pm2$ & $0.1571^{+0.0698}_{-0.0349}$ & 3 & 0.10 & 5 (9.36)\\
  130427A & 6 & 0.338 & $162.83\pm1.36$ & $808.9^{+49.6}_{-56.5}$ & $1577^{+97}_{-110}$ & $0.0663\pm0.0052$ & 3 & 0.87 & 5 (24.3)\\
   \noalign{\smallskip}\hline
  101225A & 2, 5      & 0.847 & 7000  & 240 & 100  & $>$0.21  &3& $>$0.50   & 50 (192.4)\\
  111209A &1, 2, 3, 7, 8 & 0.677 & 13000 & 570 & 5130 &$\sim$0.40&3& $\sim$3.00& 50 (1778)\\
  121027A & 2, 3, 5   & 1.773 & 6000  & 150 & 1400 & $>$0.17  &3& $>$0.76   & 50 (229.2)\\
  \hline
  %\multicolumn{9}{1}{}
\end{tabular*}\\
Notes: The duration are just $T_{90}$ for SGRBs and LGRBs. However, for ULGRBs, the duration of ULGRBs are defined as the central engine activity timescales. For SGRBs in the context of NS+BH (NS+NS) scenario, the BH mass $m$ and the upper-limit of $m_{\rm disk}$ are taken as 7 (3) and 0.5 (0.2), respectively. The BH spin $a_*$ are 0.5 for SGRBs and LGRBS, and 0.9 for LGRBs and ULGRBs. The upper-limit of $m_{\rm disk}$ are adopted as 5 for LGRBs and 50 for ULGRBs. The NDAF model without a boundary torque can not account for these GRBs, since it requires an unusually large disk mass $m_{\rm disk}$ as indicated in the brackets.
\\
Refernces --- (1) \cite{Gendre13}; (2) \cite{Levan15} and references therein; (3) \cite{Levan14} and references therein; (4) \cite{Liu15} and references therein; (5) \cite{NKSN13}; (6) \cite{Song16} and references therein; (7) \cite{Stratta13}; (8) \cite{Virgili13} and references therein.
\end{minipage}
\end{table*}

\section{Interpreting the Variability of Prompt Emission and the Followed Steep Decay Phase in X-ray Afterglow}
GRBs present remarkable time variability in their prompt emission and a steep decay phase followed by a shallow decay phase (or plateau) in  their X-ray afterglow lightcurves. The steep decay and the plateau phases are often thought to be the result of shutting-off and reactivation of the central engine, respectively. In this section, we try to interpret the variability timescale with viscous instability timescale, which is induced by the inner boundary torque. In the meanwhile, we proposed an alternative scenario for the steep decay based on the instability analysis discussed above. It's worthwhile to mention that the similar connection between the time variability and the disk instability has ever been proposed in previous works for some X-ray binaries and AGN(e.g. \citealt{Matsu89, Honma92, Ohsuga06, Ohsuga07, Oda09}).

\begin{figure}[h!tp]
\centering
\includegraphics[width=80mm]{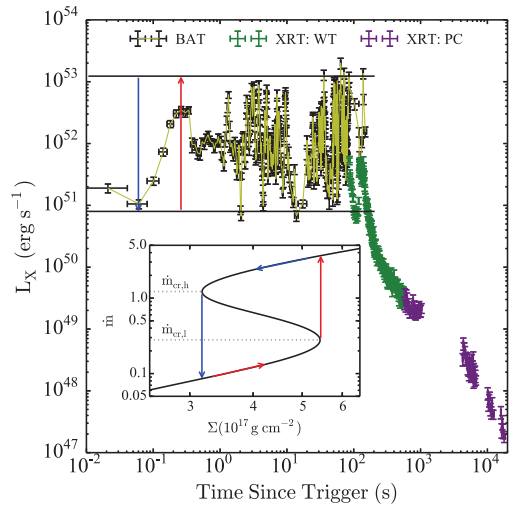}
\caption{The lightcurve in the 0.3-10 keV energy band of GRB080607. The mini panel in the figure shows the S-shape $\dot m-\Sigma$ curve at radius $r=3r_g$ which is used in the viscous instability analysis, the other parameters are $\eta=3$, $m=7$, $a_*=0.9$, $\alpha=0.1$ and $\beta=0$. The limit cycle marked by the red and blue arrows corresponds to an individual pulse in the the prompt emission stage, while the last time of the accreting rate's drop from a high one ($3.7 M_\odot\ \mathrm{s^{-1}}$) to a low one ($0.085 M_\odot\ \mathrm{s^{-1}}$) is presented as the steep decay phase in the early X-ray afterglow, in order to compare with the observed data, the two solid horizontal lines are resulted from multiplying 0.06 to the high neutrino annihilation luminosity $2.05\times 10^{54}\ \mathrm{erg\ s^{-1}}$ and the low one $1.33 \times 10^{52}\  \mathrm{erg\ s^{-1}}$, corresponding to the high and low accreting rate separately.}
\label{Fig8}
\end{figure}

We focus here on LGRBs, which are believed to be the results of the collapse of a massive star. An accretion disk will form after the collapse. As argued by \cite{KNJ08}, the mass feeding rate at the outer edge of the disk $\dot{m}_\mathrm{acc}$ decreases with time. For a fixed disk radius in the unstable region of nztNDAF, there are two critical accretion rates, i.e., a lower one $\dot{m}_{cr,l}$ (e.g., the lower turning-point in the S-curve of figure \ref{Fig8}) and a high one $\dot{m}_{cr,h}$ (e.g., the top turning-point in the S-curve of figure \ref{Fig8}). The flow is unstable when $\dot{m}_{cr,l}< \dot{m}_\mathrm{acc} < \dot{m}_{cr,h}$. Once $\dot{m}_\mathrm{acc}$ reduces to the value below the critical rate $\dot{m}_{cr,h}$, the flow quickly switches to a low accretion rate state($\dot{m}<\dot{m}_{cr,l}$). If the mass feeding rate $\dot{m}_\mathrm{acc}$ at outer edge is greater than $\dot{m}_{cr,l}$, the disk accretion rate will increase gradually until reaching $\dot{m}_{cr,l}$. The disk will then jump to the high accretion rate state with $\dot{m}>\dot{m}_{cr,h}$. If $\dot{m}_{cr,l}< \dot{m}_\mathrm{acc} < \dot{m}_{cr,h}$, the flow behaviors in a limit cycle pattern. A sequence of such cycles make up a series of individual pulses observed in the prompt emission. On the other hand, once the mass feeding rate $\dot{m}_\mathrm{acc}$ is significantly decreased to $\dot{m}_\mathrm{acc} < \dot{m}_{cr,l}$, the disk drops to the low accretion rate state but can not jump back again. These features can explain the variability in prompt emission and the followed steep decay.

As shown in figure \ref{Fig8}, we take GRB080607 (the redshift z$\sim$3.04) as an example, in which we use a nztNDAF model with $\eta = 3$. The viscous instability is triggered once the accretion rate decreases to $1.22\ M_\odot\ \mathrm{s^{-1}}$, and then it oscillates between $3.7\ M_\odot\ \mathrm{s^{-1}}$ and $0.085\ M_\odot\ \mathrm{s^{-1}}$, resulting in the change of $L_{\nu\bar{\nu}}$ between $2.05\times 10^{54}\ \mathrm{erg\ s^{-1}}$ and $1.33 \times 10^{52}\  \mathrm{erg\ s^{-1}}$. The oscillating timescale is around 100 ms (strictly speaking, the instability timescale is evaluated as viscous timescale $t_\mathrm{vis}$ which is between 6 ms and 33 ms corresponding to the unstable domain of the mass accretion rate $1.22\sim0.35\ M_\odot\ s^{-1}$, this numerical value of the viscous timescale is consistent with that derived from analytical solution (see Appendix C for more details). The observed variation timescale can be estimated as $(1+z) t_\mathrm{vis}\sim 24-133\ \mathrm{ms}$). A series of limit cycle produces a series of individual pulse, which can explain the variability in prompt emission. Finally, as the feeding rate evolves to $\dot{m}_{\rm fb} < \dot{m}_{cr,l}$,  there will be a sharp drop in luminosity, which can be considered as an natural interpretation for the steep decay phase.
    
It's worth mentioning that the value of $\eta=3$ seems rather large compared to most of the $\eta-$values in Table 1. According to Appendix A, this value is still acceptable. We use such a large $\eta$ to reproduce a significant variation in lightcurve. However, the emission originates from jet instead of disk. The amplitude of the variability may be modulated by various mechanisms, such as magnetic dissipation in the jet (e.g., \citealt{Deng15, Deng16}), precession and relativistic boosting effect (e.g., \citealt{PZLL99, Lei07, Lei15}) and so on. If we relax the constraint from the oscillation amplitude and only keep the requirement of timescale, then a much smaller $\eta$ still works. For example, if we take $\eta=1$, an accretion rate of $\dot m=0.5$ can also trigger the instability, in such case, the neutrino annihilation luminosity is about $~10^{53}\ \mathrm{erg\ s^{-1}}$, and the instability timescale is about $(1+z)*200\ \text{ms}=800\ \text{ms}$, which are still consistent with the observations. To trigger an unstable NDAF,  $\eta$ can not be too small $\eta$, otherwise it may require an extremely large $\dot m$ and therefore an unreasonable large disk mass \citep{Liu15}. So, we don't expect this model can explain all the GRBs.

\section{The Gravitational Wave Radiation from a Precessing NDAF with Boundary Torque}
As catastrophic explosion events, GRBs are believed to be promising sources of gravitational waves which might be detected by current and future detectors such as LIGO, VIRGO, DECIGO, LISA and etc. The accretion disks in GRBs may precess with time \citep{BYF96, PZLL99, RRS06, RRC10, Lei07, Liu10b}. For BH-NS binary system, if the spin axis of the BH is misaligned with the angular momentum of the binary system, the accretion disk formed after the merge would precess. For collapsar model, a inclined disk is also possible after an asymmetrical collapse. In both cases, the BH can force the misaligned disk around it to precess via Bardeen-Petterson effect \citep{BP75}. Precession also exist in the tidal disruption events (TDEs, e.g. \citealt{SL12,Lei13b}) and the active galactic nuclei (AGN, e.g. \citealt{Wu13}). \citet{Sun12} suggested that the gravitational waves from a precessing NDAF disk might be detected by DECIGO and BBO, which supplies a new probability to carry out multi-messager detection for GRBs.

The $+$ (plus) and $\times$ (cross) polarization of the gravitational wave are given as \citep{ZS79,Maggiore08}

\begin{subequations}
\begin{align}
&\begin{aligned}
h_+(t)=&h_0\sin2\theta\cos(\Omega t)\sin \iota \cos \iota +\\&2h_0 \sin^2\theta\cos(2\Omega t)(1+\cos^2 \iota),
\end{aligned}\\
&\begin{aligned}
h_\times(t)=&h_0\sin2\theta\sin(\Omega t)\sin \iota+\\&4h_0\sin^2\theta \sin(2\Omega t)\cos \iota,
\end{aligned}
\end{align}
\label{eq:20}
\end{subequations}

\noindent where

\begin{equation}
h_0=-\frac{G}{c^4}\frac{(I_3-I_1)\Omega^2}{d} \label{eq:21}
\end{equation}

\noindent In which, $d$ is the distance of the GRB, $\theta$ is the angle between the jet and the spin axis of the BH, $\iota$ denotes the LOS (line of sight) and BH spin axis. $I_1$, $I_2$, $I_3$ are the eigenvalues of the rotary inertia tensor of the inner processing part of the disk, they separatively denote the inertia along the principal axis $X$, $Y$, $Z$ and can be expressed as

\begin{subequations}
\begin{align}
&\begin{aligned}
I_1= I_2&= \int_{r<{r_{\rm p}}} \rho (x^2+z^2) dxdydz\\&=\pi\int_{r_\text{ms}}^{r_\text{p}}\Sigma r(r^2+2H^2)dr,
\end{aligned}\\
&\begin{aligned}
I_3 = \int_{r<{r_{\rm p}}} \rho (x^2+y^2) dxdydz =2\pi \int_{r_\text{ms}}^{r_\text{p}}\Sigma r^3 dr.
\end{aligned}
\end{align}
\label{eq:22}
\end{subequations}

The precession angular velocity of the accretion disk $\Omega$ is expressed as \citep{SBH80,Lu90}
\begin{equation}
\Omega=\frac{2G J_*}{c^2 r_{\rm{p}}^3}, \label{eq:23}
\end{equation}
here $J_*=G M^2 a_*/c$ denotes the BH angular moment. $r_{\rm{p}}$ is a critical radius, which is determined by equating the angular moment of this inner part with that of the BH, i.e. $J|_{r_{\rm p}}=J_*$. A typical angular momentum of the disk is $J=2\pi r^3 \Sigma v_{\phi}$, where $v_{\phi} = r\Omega_D$ is the angular velocity of disk  \citep{SBH80, Sun12}.

During GRB, the accretion disk can only exist for several to tens of seconds, suggesting that the gravitational radiation should appear as a gravitational wave burst. The gravitational waveform should be expressed as \citep{Maggiore08}

\begin{equation}
h(t)=[h_+(t)+h_{\times}(t)]\exp{\left(-\frac{t^2}{2\delta^2}\right)},
\label{eq:24}
\end{equation}
\noindent where $\delta$ is the duration of the GRBs. We take $\delta \simeq 20 s$ as a typical value in the calculations. The root-sum-square (rss) amplitude of the gravitational wave is used to estimate the detectability \citep{AAA08, Maggiore08},

\begin{equation}
h_\text{rss}=\sqrt{\int_{-\infty}^\infty\left(h_+^2(t)+h_\times^2(t)\right)dt}.
\label{eq:25}
\end{equation}

The gravitational radiating power of the precession disk is expressed as
\begin{equation}
P_\text{GW}=\frac{2G}{5c^5}(I_1-I_3)^2\Omega^6\sin^2\theta(1+15\sin^2\theta).
\label{eq:26}
\end{equation}

We recalculate the GW strength and frequency from the NDAF according to the same procedure of \citet{Sun12}. In our calculations, we adopt one-zone approximation for the vertical structure of the disk, which is embodied in the deducing process of equation (\ref{eq:22}). In this section we aim to investigate the significance of the effects of the inner edge torque on gravitational wave radiation.

\begin{figure}[!ht]
\centering
\includegraphics[height=40mm]{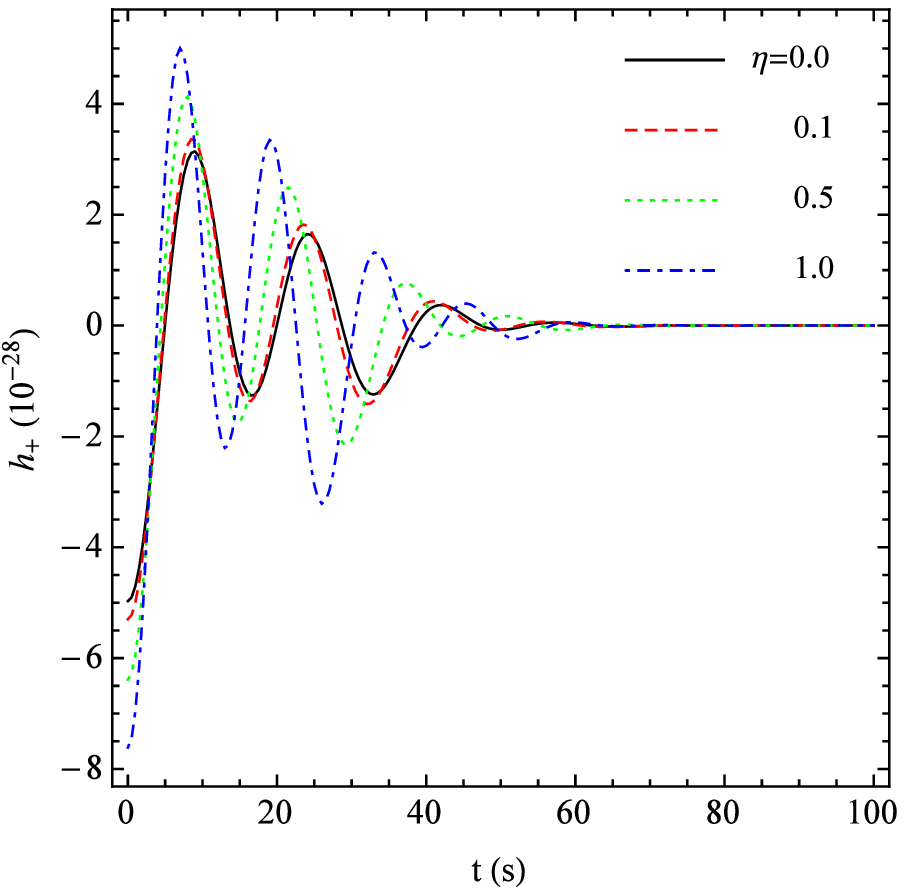}
\includegraphics[height=40mm]{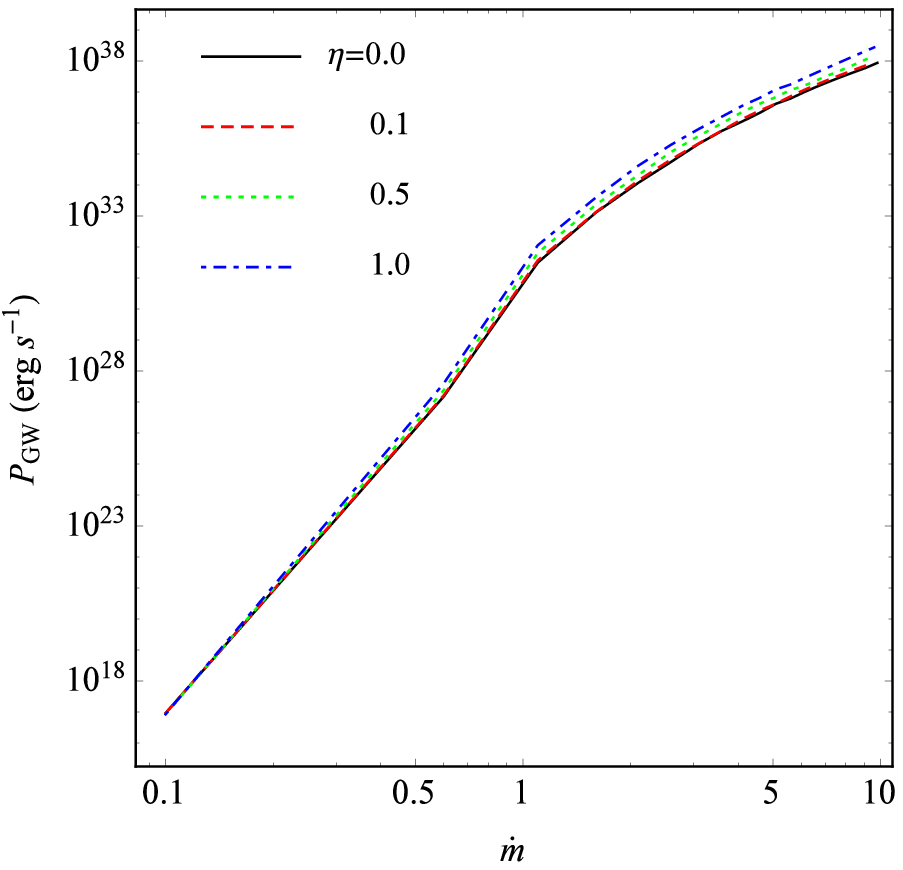}\\
\includegraphics[height=40mm]{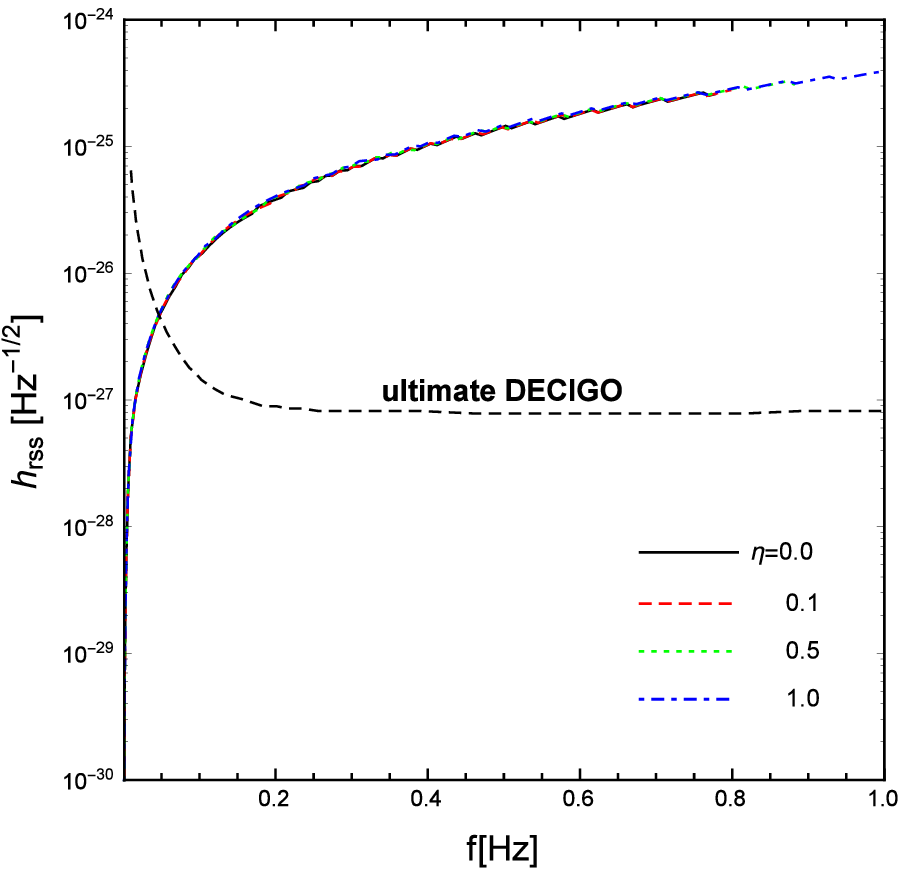}
\caption{The effects of the inner edge torque on the gravitational wave radiated from a precessing disk. The upper left panel shows one of the polarization mode of the gravitational wave from a precessing disk with different inner edge parameter $\eta$; the upper right panel shows the power of gravitational radiation versus accretion rate with different inner edge parameter $\eta$, the lower panel shows the root-sum-square amplitude versus the frequency of the gravitational wave with different inner edge parameter $\eta$. Other parameters are $m=7$, $\alpha=0.1$; $\beta=0.0$, $\dot{m}=1.0$, $\theta=20^\circ$, $\iota=20^\circ$, and $d=1 \text{Mpc}$.}
\label{Fig9}
\end{figure}

Figure \ref{Fig9} shows the gravitational wave from a precessing NDAF disk with different inner edge torque. Although the inner edge torque significantly changes the disk properties, it has little effect on the gravitational wave radiation from the precessing disk. This result is reasonable, since the boundary torque only weakly change the mass distribution of the outer disk. With the increase of $g_{\rm ms}$, the gravitational wave's frequency, amplitude and radiation power slightly represents the increment.

%%%%%%%%%%%%%%%%%%%%%%%%%%%%%%%%%%%%%%%%%%%%%%%%%%%%%%%%%%%%%%%%%%%%%%%%%%%%%%%%%%%%%%%%%%%
% CONCLUSIONS
%%%%%%%%%%%%%%%%%%%%%%%%%%%%%%%%%%%%%%%%%%%%%%%%%%%%%%%%%%%%%%%%%%%%%%%%%%%%%%%%%%%%%%%%%%%

\section{Conclusions and Discussions}
\label{sect:Discussion}
We revise the NDAF model by including a boundary stress. Based on numerical and analytical solutions, we study the properties of nztNDAF. The disk becomes much hotter and denser due to the non-zero boundary torque. The properties in inner region are significantly different from those of NDAF. As a result, we find that the disk becomes unstable if $\dot{m}$ and $\eta$ are great enough (e.g., when the parameter $\eta>0.45$ for $\dot{m}=1.0$). The neutrino annihilation luminosity is greatly enhanced by the boundary stress. The luminosity of nztNDAF varies from $7.8\times10^{48}\rm{erg\, s^{-1}}$ to $2.4\times10^{54}\rm{erg\, s^{-1}}$ for $0.01<\dot{m}<10$ with $\eta=1$, which is much greater than NDAF (its range is from $2.3\times10^{45}\rm{erg\, s^{-1}}$ to $4.6\times10^{53}\rm{erg\, s^{-1}}$).

We then apply nztNDAF model to GRBs. For some bright short GRBs and powerful long GRBs,  NDAF model is challenged when interpreting the limited mass of the accretion disk after the compact binaries coalescence or massive collapsar \citep{Liu15, Song16}. For the same reason, NDAF is not expected to explain ULGRBs. However, in this paper we argued that the NDAF could still be a feasible model for those issued GRBs, as long as the inner boundary torque is considered. In addition, we extend the method of \cite{Liu15} to ultra-long GRBs and find that the nztNDAF model is also suitable under the frame of BSG-progenitor.

Viscous instability may occur nztNDAF in the inner region. When it happens, the disk will transit between two stable brunch with different accretion rates, leading to a variable jet luminosity. The timescale for the instability is about 10 ms (estimated by the viscous timescale at the inner disk). These results can explain the variability in GRB lightcurves. The steep decay following the prompt emission occurs when the mass feeding rate at the outer edge of the disk reduces to a value lower than a critical accretion rate. Finally, we find that the effects of boundary torque on gravitational wave can be ignored.

In this work, we describe the boundary torque with a parameter $\eta$. The properties of nztNDAF strongly depend on the value of $\eta$. However, there is no characteristic or ``natural'' magnitude that one can select for the torque \citep{ZNMM05}. Numerical simulation performed by \citet{PSM12} indicated the stress at the inner edge to be directly proportional to the disk thickness. They then argued that zero-stress boundary condition is valid for thin disk in the limit $h\rightarrow 0$. However, the GRMHD simulations by \citet{Noble10} with different thickness found a large stress at the inner edge even in the limit of vanishing disk opening angle $h\rightarrow 0$. The GRMHD simulations by \citet{K05} and \citet{B08} also found that the torque can reach a very high value in the plunging region. Due to these uncertainties, we take $\eta$ as a free parameter in the calculations. In these works, the magnetic stress in the plunging region is likely the origin of the non-zero torque at inner disk edge. For simplicity, we roughly take the magnetic coupling torque exacted by BH as the upper limit of such boundary torque \citet{Lei09}. Nonetheless, there are two differences at least between the nztNDAF model in this work and the MCNDAF model in \citet{Lei09}. Firstly, the MCNDAF essentially adopted the zero stress assumption at the inner edge of the disk, and the MC torque is a resultant effect of the magnetic stresses differentially distributed in a limited disk region which is coupled with the BH by ordered large scale magnetic field lines (see the integrated angular momentum equation (18) in \citet{Lei09}). While in the nztNDAF model, the non-zero stress is just exerted on the inner edge rather than any location else, this inner edge stress originates from the angular momentum transport between the plumping region and the disk through magnetized turbulence. Secondly, for the same magnitude of the two different types of extra torque, the extra viscous heating rate in the inner region caused by the inner edge torque in nztNDAF is much higher than that of the MCNDAF, consequently, the structure near the inner edge of the nztNDAF will be changed with a more significant extent than MCNDAF. Those arguments deserve further studying by GRMHD simulation.

For simplicity, we just consider radial direction in this work, with adopting the one-zone approximation in the vertical direction. Our current results show that the inner side of the disk will expand a lot due to the extra heating by the non-zero torque (Figure 1d). Especially when $\eta \gtrsim 1$, the ratio of height to radius $h/r$ can approach unity. The larger disk height will aggravate the extent to which the neutrinos are trapping in the disk, this is one of the reason why the innermost disk become advection cooling dominated. In addition, the increment of the disk height due to the inner edge torque indicates the necessity of the further consideration of the vertical structure. According to \cite{G07}, when $h/r \gtrsim 0.2$, the H\={o}shi form of the gravitational potential \citep{H77} used for deriving the vertical static equilibrium can not be satisfactory any more. There are a number of works on the vertical structure of NDAF (e.g., \citealt{Saw03, Liu10, Liu12a, Liu13, Liu14, Liu15a, PY12}). We may further explore the effects of the inner edge torque by considering the vertical structure in future.

Another widely discussed GRB central engine model is Blandford-Znajek mechanism \citep{BZ77, Lee00, Lei13a}. \citet{Lei13a} studied the baryon loading in NDAF and Blandford-Znajek jets. It is found that Blandford-Znajek mechanism can produce ``clean'' jet. NDAF-driven ``fireball'' is typically too ``dirtier'' to account for GRBs. In our nztNDAF model, however, the existence of the magnetic filed may help to suppress baryons from disk, and thus lead to a clean central engine.

% \section{Acknowledgments}

\begin{acknowledgements}
We thank Bing Zhang and Hui Li for constructive suggestions. We also thank the anonymous referee for his/her valuable comments and constructive suggestions. This work is supported by National Basic Research Program (``973'' Program)
of China under grant No. 2014CB845800, National Natural Science Foundation
of China under grants U1431124, 11361140349 (China-Israel jointed program).
\end{acknowledgements}

%%%%%%%%%%%%%%%%%%%%%%%%%%%%%%%%%%%%%%%%%%%%%%%%%%%%%%%%%%%%%%%%%%%%%%%%%%%%%%%%%%%%%%%%%%%
% BIBLIOGRAPHY
%%%%%%%%%%%%%%%%%%%%%%%%%%%%%%%%%%%%%%%%%%%%%%%%%%%%%%%%%%%%%%%%%%%%%%%%%%%%%%%%%%%%%%%%%%%

%----------------------------------------------------------------------------------------

%

%***************************************************************

%-------------------Appendix-----------------------
\newpage
\appendix
\section{Appendix A: Constrains on the Inner Boundary Torque}\label{appendix A}
In nztNDAF, the magnetic stress in the plunging region is likely the origin of the non-zero torque at inner disk edge. As we known, the BH can exert a torque on the disk through the magnetic coupling mechanism (MC) and transfer energy and angular momentum to disk \citep{B99, van99, LP00, Li02, Wang02, Lei09}. Based on this scenario, we can put an upper limit for $\eta$.

The magnetic field at the BH horizon could be derived by using the equipartition relation as follows \citep{M05}:
 \begin{equation}
  \frac{B_\mathrm{H}^2}{8\pi}=\rho_{0,\mathrm{disk}}c^2,
  \label{eq:A1}
 \end{equation}
 where $\rho_{0,\mathrm{disk}}\equiv\dot{M}t_g/r_g^3$, $t_g=GM/c^3$, and $r_g=GM/c^2$.

 The MC torque is expressed as\citep{Wang02}:
 \begin{equation}
 T_\mathrm{MC}/T_0=f(a_*;n)=4a_*(1+q)\int_0^{\pi/2} \frac{(1-\beta)\sin^3\theta \mathrm{d}\theta}{2-(1-q)\sin^2\theta},
 \label{eq:A2}
 \end{equation}
 in which $q\equiv \sqrt{1-a_*^2}$, $T_0=B_\mathrm{H}^2 \left(\frac{GM}{c^2}\right)^3$. $\beta$ denotes the ration of the angular velocity of the disk to that of the BH horizon, which is defined by
 \begin{equation}
 \beta\equiv\Omega_\mathrm{D}/\Omega_H,
 \label{eq:A3}
 \end{equation}
where $\Omega_\mathrm{D}=c^3/GM(\chi^3+a_*)$, $\Omega_H=a_*c^3/2GM(1+q)$, $\chi\equiv\sqrt{r/r_g}$.
Combining eq.(\ref{eq:A1}) and eq.(\ref{eq:A2}), one gets the MC torque
\begin{equation}
T_\mathrm{MC}=\frac{8\pi GM\dot{M}}{c}f(a_*;n).
\label{eq:A4}
\end{equation}

Based on the conservation of magnetic flux, \cite{Wang03} proposed the mapping relation between the angular coordinate $\theta$ on the horizon and the radial coordinate $\xi$ (defined as $\xi\equiv r/r_\mathrm{ms}$):
\begin{equation}
\cos\theta=\int_1^\xi \Theta(a_*;\xi,n) \mathrm{d}\xi,
\label{eq:A5}
\end{equation}

\begin{equation}
\Theta(a_*;\xi,n)=\frac{\chi_\mathrm{ms}^4\xi^{1-n}}{2(1+q)}\sqrt{\frac{1+a_*^2\chi_\mathrm{ms}^{-4}\xi^{-2}+
2a_*^2\chi_\mathrm{ms}^{-6}\xi^{-3}}{1-2\chi_\mathrm{ms}^{-2}\xi^{-1}+a_*^2\chi_\mathrm{ms}^{-4}\xi^{-2}}}.
\label{eq:A6}
\end{equation}

On the other hand, we have introduced a parameter $\eta$ to reflect the strength of the inner boundary torque of the disk, we rewrite it as follow:
\begin{equation}
T_{MC}=\eta \dot{M} L_\mathrm{ms}.
\label{eq:A7}
\end{equation}

Combining eq.(\ref{eq:A4}) and eq.(\ref{eq:A7}), we have
\begin{equation}
\eta=8\pi f(a_*;n)/\zeta(a_*),
\label{eq:A8}
\end{equation}
where $\zeta(a_*)=L_{\rm ms} c/GM$.

\begin{figure*}[h!tp]
\centering
\includegraphics[width=80mm]{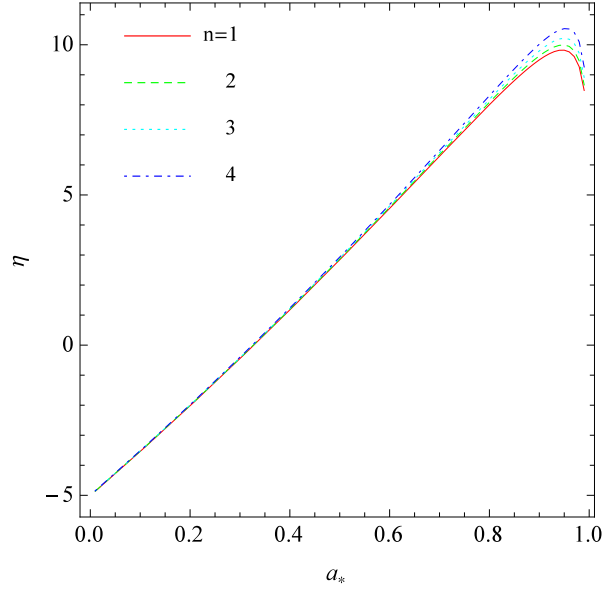}
\caption{$\eta$ versus the BH spin $a_*$ under different magnetic field configuration parameter $n$.}
\label{MC_torque}
\end{figure*}

Figure \ref{MC_torque} shows the variation of $\eta$ versus the BH spin $a_*$ for different magnetic field configuration parameter $n$. We find that $\eta$ rapidly increase with $a_*$, but is not very sensitive to $n$.

\section{Appendix B: The Angular Momentum Conservation Equation}\label{appendix B}
The differential equation of the conservation of angular momentum is given by \cite{RH95} as
\begin{equation}
\frac{\partial \tau_{r\phi}}{\partial r}=-\frac{2}{rA}\left(1-\frac{M}{r}\right)\tau_{r\phi}-\frac{\sqrt{M}}{2r^{3/2}}\frac{E}{AB}r\rho u^r,
\label{eq:B1}
\end{equation}
here we use natural units $G=c=1$, $u^r$ is the radial component of the four velocity. Integrating over the disk height, equation (\ref{eq:B1}) becomes

\begin{equation}
\frac{\mathrm{d}}{\mathrm{d}r}\left(2\pi\int_{-\infty}^\infty\tau_{r\phi}\mathrm{d}z\right)=
-\frac{2}{rA}\left(1-\frac{M}{r}\right)2\pi\int_{-\infty}^\infty\tau_{r\phi}\mathrm{d}z
-\frac{\sqrt{M}}{2r^{3/2}}\frac{E}{AB}2\pi r u^r \int_{-\infty}^\infty\rho  \mathrm{d}z.
\label{eq:B2}
\end{equation}

\noindent The continuity equation is

\begin{equation}
\dot{M}=-2\pi r u^r\int_{-\infty}^\infty\rho \mathrm{d}z =-2\pi r (2h) \rho u^r,
\label{eq:B3}
\end{equation}

\noindent Inserting equation (\ref{eq:B3}) into equation (\ref{eq:B2}), we have

\begin{equation}
\frac{\mathrm{d}\Lambda}{\mathrm{d}r}+P(r)\Lambda=Q(r),
\label{eq:B5}
\end{equation}
where
\begin{equation}
\begin{aligned}
&\Lambda=2\pi\int_{-\infty}^\infty\tau_{r\phi}\mathrm{d}z \\
&P(r)=\frac{2}{rA}\left(1-\frac{M}{r}\right) \\
&Q(r)=\frac{\sqrt{M}}{2r^{3/2}}\frac{E}{AB}\dot{M}.
\end{aligned}
\label{eq:B6}
\end{equation}
Note that $\Lambda$ is related to the torque $g$ by $g=r^2 \Lambda$.

To solve equation (\ref{eq:B5}), we introduce a function as follow,

\begin{equation}
\xi(r)=e^{\int P(r) \mathrm{d}r}.
\label{eq:B7}
\end{equation}

\noindent Substituting the expression of $P(r)$ in equation (\ref{eq:B6}), equation (\ref{eq:B7}) can be rewritten as,

\begin{equation}
\xi(r)=r^2\left(1-\frac{2M}{r}+\frac{a^2}{r}\right)=r^2 A,
\label{eq:B8}
\end{equation}

\noindent Multiplying both sides of equation (\ref{eq:B5}) by the factor $\xi(r)$, we have

\begin{equation}
\frac{\mathrm{d}\xi \Lambda}{\mathrm{d}r}=\xi(r) Q(r).
\label{eq:B9}
\end{equation}

\noindent Integrating equation (\ref{eq:B9}) from $r_{\text{ms}}$ to $r$, one gets

\begin{equation}
\xi(r) \Lambda(r)-\xi(r_{\rm ms}) \Lambda(r_{\rm ms})=\int_{r_{\rm ms}}^r \xi(r) Q(r)\mathrm{d}r.
\label{eq:B10}
\end{equation}
It can be reduced to

\begin{equation}
\Lambda(r)=\dot{M}\sqrt{\frac{M}{r^3}}\frac{D}{A}+\frac{r_{\rm ms}^2 A_{\rm ms}}{r^2 A} \Lambda_{\rm ms},
\label{eq:B12}
\end{equation}
where $D=\frac{1}{2\sqrt{r}}\int_{r_{\rm ms}}^r \frac{E}{\sqrt{r}B}\mathrm{d}r$.

\noindent Dividing both sides of equation (\ref{eq:B12}) by $4\pi h$, we have

\begin{equation}
\tau_{r\phi}(r)=\frac{\dot{M}}{4\pi h}\sqrt{\frac{M}{r^3}}\frac{D}{A}+\frac{r_{\rm ms}^2 A_{\rm ms}}{r^2 A}\tau_{r\phi}(r_{\rm ms}).
\label{eq:B13}
\end{equation}

\noindent Multiplying equation (\ref{eq:B13}) by $4\pi r^2 h$ and using $g=4\pi r^2 h\tau _{r\phi}$, we have

\begin{equation}
g(r)=\dot{M}r^2\sqrt{\frac{M}{r^3}}\frac{D}{A}+\frac{A_{\rm ms}}{A}g_{\rm ms}.
\label{eq:B14}
\end{equation}

\section{Appendix C: Analytical Solutions of NDAF with a Non-Zero Boundary Torque}\label{appendix C}
We dedicate this section to obtain analytical solutions of the nztNDAF under the indication of the previous numerical results. These analytical solutions are helpful to understand the characteristics of nztNDAF.

For convenience of illustration, here we rewrite some dynamical equations and other more details. Combining eq.(\ref{eq:3}), eq.(\ref{eq:9}) and eq.(\ref{eq:10}) one gets an expression for the total pressure as
\begin{equation}
P=\left[\frac{GM\dot{M}\rho^{1/2}}{4\pi \alpha r^3}\frac{C\mathscr{D}}{A^2}\right]^{2/3},
\label{eq:C1}
\end{equation}
 where $\mathscr{D}\equiv D\left(1+\eta\frac{L_\mathrm{ms}}{r^2\Omega_\mathrm{k}}\frac{A_\mathrm{ms}}{D}\right)$. Note that the dynamical impact of the inner edge (i.e., $\eta$) is embedded in symbol $\mathscr{D}$.

As stated before, the total pressure $P$ is composed of several components such as the gas pressure, the radiation pressure, degeneracy pressure, neutrino pressure and magnetic pressure, which is expressed as
\begin{equation}
P=P_{\rm{gas}}+P_{\rm{rad}}+P_{\rm{deg}}+P_{\rm{\nu}}+P_{\rm{B}},
\label{eq:C2}
\end{equation}
\noindent each term on the right side of the above expression is separatively given bellow,
\begin{subequations}
\begin{align}
&P_\text{rad}=\frac{11}{12}a T^4, \\
&P_\text{gas}=\frac{\rho k T}{m_p}\left(\frac{1+3X_{\text{nuc}}}{4}\right),\\
&P_\text{deg}=\frac{2 \pi h c}{3}\left(\frac{3}{8 \pi m_p}\right)^{4/3}\left(\frac{\rho}{\mu_e}\right)^{4/3},\\
&P_\nu=\frac{1}{3}u_\nu,\\
&P_{\rm{B}}=\beta P.
\end{align}
\label{eq:C3}
\end{subequations}

\noindent Equation (\ref{eq:C2}) attached with equation (\ref{eq:C3}) is known as the equation of state of the accretion flow.

The energy equation is expressed as,
\begin{equation}
Q_{\rm{vis}}^+ =Q_{\rm{\nu}}^- +Q_{\rm{photo}}^- +Q_{\rm{adv}}^-,
\label{eq:C4}
\end{equation}
\noindent in which the viscous heating rate is

\begin{equation}
Q_{\rm{vis}}^+=\frac{3GM\dot{M}}{8\pi r^{3}}\frac{\mathscr{D}}{B}.
\label{eq:C5}
\end{equation}

\noindent The cooling rate due to neutrino losses $Q_\nu^-$, photodisintegration $Q_{\rm photo}^-$, and advection $Q_{\rm adv}^-$ are expressed as,

\begin{subequations}
\begin{align}
&Q_{\nu}^- = \sum_i \frac{(7/8)\sigma T^4}{(3/4)(\tau_{\nu_i}+1/\sqrt{3}+1/3\tau_{a,\nu_i})}\\
&Q_{\text{photo}}^{-}\simeq 10^{29} \rho_{10} v_r h \frac{\text{d}X_\mathrm{nuc}}{\text{d}r},\\
&\begin{aligned}
Q_\text{adv}^{-}=\Sigma v_r T \frac{\text{d}s}{\text{d}r} \ \ \ \ \ \ \ \ \ \ \ \ \ \ \ \ \ \ \ \ \ \ \ \ \ \ \ \ \ \ \ \ \ \ \ \ \ \ \ \ \ \\ \simeq \xi v_r \frac{H}{r}T \left(\frac{11}{3} a T^3 +\frac{3}{2} \frac{\rho k}{m_p}\frac{1+X_\text{nuc}}{4}+{\frac{4}{3}\frac{u_\nu}{T}}\right) \ \ \\
=\xi \frac{\dot M}{4\pi r^2 \rho}\left(\frac{11}{3} a T^4+\frac{3}{2}\frac{\rho k T}{m_p}\frac{1+X_\text{nuc}}{4} +\frac{4}{3}u_\nu\right),
\end{aligned}
\end{align}
\label{eq:C6}
\end{subequations}

\noindent where $\tau_{\nu_i}=\tau_{a,\nu_i}+\tau_{s,\nu_i}$ is the sum of the absorption and scattering optical depth for each neutrino flavor ($\nu_e, \nu_\mu, \nu_\tau$). The absorption optical depth include the contributions from the interaction of neutrinos with one another $\tau_{a,\nu_i \bar{\nu}_i}$, the absorption onto protons or onto neutrons $\tau_{a,eN}$. The expressions for these optical depth are,
\begin{subequations}
\begin{align}
&\tau_{a,\nu_i \bar{\nu}_i} \approx 2.5\times10^{-7} T_{11}^5 h\\
&\tau_{a,eN}\approx4.5\times10^{-7} T_{11}^2 X_\mathrm{nuc} \rho_{10} h\\
&\tau_{s,\nu_i}\approx2.7\times10^{-7} T_{11}^2 \rho_{10} h
\end{align}
\label{eq:C7}
\end{subequations}

Inspecting our numerical solutions, we found that the whole disk can be divided into several different characteristic regions, as shown in Figure \ref{sketch} and Table 2. We will obtain the analytical solutions for each region in the following subsections.

\begin{table*}
\label{table_regions}
\centering
\caption{The different regions in a hyper-accretion disk characterised by different dominant pressure component, cooling mechanism or neutrino opacity.}
\begin{tabular}{|l|c|c|c|c|c|c|}
  \hline
  \diagbox{Property}{Regions} & \tabincell{c}{(VI)\\ADAF} & \tabincell{c}{(V)\\ADAF}  & \tabincell{c}{(IV)\\unstable NDAF} & \tabincell{c}{(III)\\opaque NDAF} & \tabincell{c}{(II)\\transparent NDAF} & \tabincell{c}{(I)\\ADAF} \\
  \hline
  gas pressure dominated       &$\surd$&       &       &$\surd$&$\surd$&       \\
  radiation pressure dominated &       &$\surd$&$\surd$&       &       &$\surd$\\
  neutrino cooling dominated   &       &       &$\surd$&$\surd$&$\surd$&       \\
  advection cooling dominated  &$\surd$&$\surd$&       &       &       &$\surd$\\
  $\nu-$opaque        &       &       &$\surd$&$\surd$&       &       \\
  $\nu-$transparent         &       &       &       &       &$\surd$&       \\
  \hline
\end{tabular}
\end{table*}

\subsection{Region I --- radiation pressure dominated ADAF}
At large radii, the disk could be dominated by advection cooling since the radiation cooling timescale is much longer than the accretion timescales. The mass density in this region is relative small and the temperature is still so high that the radiation pressure dominates the accretion flow. The equation of state is $P\approx P_\mathrm{rad}+P_\mathrm{B}=(1-\beta)^{-1} P_\mathrm{rad}$, and energy conservation equation is $Q_\mathrm{vis}^+\approx Q_\mathrm{adv}^-$. Therefore, we have,
\begin{subequations}
\begin{align}
&\left(\frac{GM\dot{M}\rho^{1/2}}{4\pi \alpha r^3}\frac{C\mathscr{D}}{A^2}\right)^{2/3}=(1-\beta)^{-1}\frac{11}{12}aT^4,\\
&\frac{3GM\dot{M}}{8\pi r^{3}}\frac{\mathscr{D}}{B}=\frac{\dot M}{4\pi r^2 \rho} \frac{11}{3} a T^4,
\end{align}
\label{eq:C8}
\end{subequations}
\noindent Note that the magnetic pressure is always kept in the deducing process for the convenience to analyze the effect of parameter $\beta$. However, we take $\beta=0$ in the calculation later on due to its limited influence on the disk structure. From The above equations we can obtain analytical solution of $\rho$ and $T$. Furthermore, substituting the expressions of $\rho$ and $T$ into equations (\ref{eq:10}) and (\ref{eq:B3}), one gets the solution of disk scalar height $h$ and advection velocity $u^r$. The expressions for these parameters are,

\begin{subequations}
\begin{align}
&\rho=1.05\times10^{12} (1-\beta)^{3/2} A^{-2}B^{3/2}C\mathscr{D}^{-1/2}\alpha^{-1}m^{-2}\dot{m} R^{-3/2} \mathrm{\ \ \ g\ cm^{-3}}\\
&T=4.76\times10^{11} (1-\beta)^{3/8}  A^{-1/2}B^{1/8}C^{1/4}\mathscr{D}^{1/8}\alpha^{-1/4}m^{-1/2}\dot{m}^{1/4}R^{-5/8} \mathrm{\ \ \ K}\\
%&h=9.04\times10^4C^{-1/2}\mathscr{D}^{1/2}mR \mathrm{\ \ \ cm}\\
&H=0.61C^{-1/2} (1-\beta)^{-1/2}  \mathscr{D}^{1/2}R\\
&u^r=1.12\times10^{10} (1-\beta)^{-1}  A^2B^{-3/2}C^{-1/2}\alpha R^{-1/2} \mathrm{\ \ \ cm/s},
\end{align}
\label{eq:C9}
\end{subequations}

\noindent where $H \equiv h/r_g$ is the dimensionless disk height. The radial profile for these parameters are shown in Figures \ref{Fig:C1} and \ref{Fig:C2} (see Region I).

\subsection{Region II --- transparent NDAF}
Just inside region I (ADAF), the disk is a NDAF region since the temperature and density are high enough to ignite neutrino cooling there. The neutrino opacity is not important in this region, so it is a $\nu$-transparent NDAF. For such thin disk, cooling by pair capture on nuleons $Q^-_{\rm eN}$ should dominate over by electron-positron pair annihilation, and the neutrino cooling rate in equation (\ref{eq:C6}) can be reduced to,

\begin{equation}
Q_{\nu}^- (\approx Q_\mathrm{eN}^-) \approx \frac{(7/8)\sigma T^4}{(3/4)(1/3\tau_{a,\nu_e})}=C_1\rho T^6 X_\mathrm{nuc}h \mathrm{\ \ \ ergs\ cm^{-3}\ s^{-1}},
\label{eq:C10}
\end{equation}

\noindent  where $C_1=9.0\times10^{-43}$, and we approximately take $X_\mathrm{nuc}\sim 1$ hereafter. The pressure is dominated by gas pressure, i.e, $P\approx (1-\beta)^{-1}P_\text{gas}$, and we have,

\begin{equation}
\left(\frac{GM\dot{M}\rho^{1/2}}{4\pi \alpha r^3}\frac{C\mathscr{D}}{A^2}\right)^{2/3}=(1-\beta)^{-1}\frac{\rho k T}{m_\mathrm{p}}.
\label{eq:C11}
\end{equation}

\noindent The neutrino cooling term dominates now, therefore $Q_{\rm{vis}}^+=Q_{\rm{eN}}^-$, or,
\begin{eqnarray}
\frac{3GM\dot{M}}{8\pi r^{3}}\frac{\mathscr D}{B}
=C_1\rho T^6 h =C_1\rho T^6\sqrt{\frac{(1-\beta)^{-1}P_\text{gas}r^3}{\rho GM}}\sqrt{\frac{B}{C}}
=C_1\rho T^6\sqrt{\frac{(1-\beta)^{-1} k T r^3}{GMm_\mathrm{p}}}\sqrt{\frac{B}{C}}.
\label{eq:C12}
\end{eqnarray}

\noindent The solutions can be worked out in the same way as for region I. The expressions are collected as follows,
\begin{subequations}
\begin{align}
&\rho=2.02\times10^{14}(1-\beta)^{9/5}A^{-13/5}B^{9/20}C^{23/20}\mathscr{D}\alpha^{-13/10}m^{-17/10}\dot{m}R^{-51/20} \mathrm{\ \ \ g\ cm^{-3}},\\
&T=1.23\times10^{11}(1-\beta)^{-1/5} A^{2/5}B^{-3/10}C^{-1/10} \alpha^{1/5} m^{-1/5}R^{-3/10} \mathrm{\ \ \ K},\\
%&h=1.57\times10^{4}A^{1/5}B^{7/20}C^{-11/20}\alpha^{1/10}m^{9/10}R^{27/20} \mathrm{\ \ \ cm},\\
&H=0.11 (1-\beta)^{-3/5}A^{1/5}B^{7/20}C^{-11/20}\alpha^{1/10}m^{-1/10}R^{27/20},\\
&u^r=3.38\times10^{8}(1-\beta)^{-6/5}A^{12/5}B^{-4/5}C^{-3/5}\mathscr{D}^{-1}\alpha^{6/5}m^{-1/5}R^{1/5} \mathrm{\ \ \ cm\ s^{-1}},\\
&Q_{\nu}^-=Q_{\rm{vis}}^+=9.79\times10^{42} B^{-1}\mathscr{D}m^{-2}\dot{m}R^{-3} \mathrm{\ \ \ erg\ cm^{-2}\ s^{-1}},\\
&P_\mathrm{rad}/P=7.70\times10^{-4}(1-\beta)^{-7/5}A^{19/5}B^{-27/20}C^{-29/20}\mathscr{D}^{-1} \alpha^{19/10}m^{11/10}\dot{m}^{-1}R^{33/20},\\
%&f_\mathrm{adv}\equiv Q_\mathrm{adv}^-/Q_\mathrm{vis}^+=5.64\times10^{-3}A^{2/5}B^{7/10}C^{-1/10}\mathscr{D}^{-1}\alpha^{1/5}m^{-1/5}R^{7/10},\\
&\tau_{\nu_e}=3.44\times10^{2}(1-\beta)^{4/5}A^{-8/5}B^{1/5}C^{2/5}\mathscr{D} \alpha^{-4/5}m^{-6/5}\dot{m}R^{-9/5}.
\end{align}
\label{eq:C13}
\end{subequations}

In our analytical calculations, the transition between region II and region I is roughly determined by $P_\mathrm{rad}/P=0.2$, this value is chosen so as to minimize the gap from region I to region II. For the same reason, we take the position where $\tau_{\nu_e}=2.5$ as the transition between region II (transparent NDAF) and region III (opaque NDAF).

\subsection{Region III --- opaque NDAF}
Going further inward, the mass density and temperature gradually increase until the neutrino opacity becomes important. The disk enters region III, i.e., a $\nu$-opaque NDAF. The gas pressure still dominates the flow in this region. As the neutrino optical depth is so high, equation (\ref{eq:C6}) can be reduced to

\begin{equation}
Q_{\nu}^- = \frac{7}{3}\sum_i \frac{\sigma T^4}{\tau_{\nu_i}}\approx \frac{7}{3} \sigma T^4 \left(\frac{1}{\tau_{a,eN}+\tau_{s,\nu_e}}+\frac{1}{\tau_{s,\nu_\mu}}+\frac{1}{\tau_{s,\nu_\tau}}\right)
\label{eq:C14}
\end{equation}

\noindent Note that we ignore $\tau_{a, \nu_i, \bar{\nu}_i}$ here. Substituting the equation (\ref{eq:C7}b) and (\ref{eq:C7}c) into equation (\ref{eq:C14}), one gets

\begin{equation}
Q_{\nu}^- = 1.16\times10^{35}\rho^{-1}T^2h^{-1} \mathrm{\ \ \ ergs\ s^{-1}}.
\label{eq:C15}
\end{equation}

\noindent From the above discussions, we have the equations $P\approx(1-\beta)^{-1}P_\mathrm{gas}$ and $Q_{\rm{vis}}^+=Q_\nu^-$ in this region. They can be rewritten as,

\begin{subequations}
\begin{align}
&\left(\frac{GM\dot{M}\rho^{1/2}}{4\pi \alpha r^3}\frac{C\mathscr{D}}{A^2}\right)^{2/3}\approx (1-\beta)^{-1}\frac{\rho k T}{m_\mathrm{p}},\\
&\frac{3GM\dot{M}}{8\pi r^{3}}\frac{\mathscr D}{B}\approx1.16\times10^{35}\rho^{-1}T^2h^{-1}\approx1.16\times10^{35}\rho^{-1}T^2 \left(\frac{GMm_\mathrm{p}}{r^3(1-\beta)^{-1}kT}\frac{C}{B}\right)^{1/2}.
\end{align}
\label{eq:C16}
\end{subequations}

\noindent The solutions in this region are collected as follow,
\begin{subequations}
\begin{align}
&\rho=1.52\times10^{12}(1-\beta)A^{-1}B^{1/4}C^{3/4}\alpha^{-1/2}m^{-1/2}R^{-3/4} \mathrm{\ \ \ g\ cm^{-3}},\\
&T=3.20\times10^{12}(1-\beta)^{1/3} A^{-2/3}B^{-1/6}C^{1/6}\mathscr{D}^{2/3}\alpha^{-1/3} m^{-1}\dot{m}^{2/3}R^{-3/2} \mathrm{\ \ \ K},\\
%&h=8.01\times10^{4}A^{-1/3}B^{5/12}C^{-5/12}\mathscr{D}^{1/3}\alpha^{-1/6}m^{1/2}\dot{m}^{1/3}R^{3/4} \mathrm{\ \ \ cm},\\
&H=0.54 (1-\beta)^{-1/3}A^{-1/3}B^{5/12}C^{-5/12}\mathscr{D}^{1/3} \alpha^{-1/6}m^{-1/2}\dot{m}^{1/3}R^{3/4},\\
&u^r=8.82\times10^{9}(1-\beta)^{-2/3}A^{4/3}B^{-2/3}C^{-1/3}\mathscr{D}^{-1/3} \alpha^{2/3}m^{-1}\dot{m}^{2/3}R^{-1} \mathrm{\ \ \ cm\ s^{-1}},\\
&Q_{\nu}^-=Q_{\rm{vis}}^+=9.79\times10^{42} B^{-1}\mathscr{D}m^{-2}\dot{m}R^{-3} \mathrm{\ \ \ erg\ cm^{-2}\ s^{-1}},\\
&P_\mathrm{rad}/P=6.61\times10^{2}(1-\beta)A^{-1}B^{-3/4}C^{-1/4}\mathscr{D}^{2} \alpha^{-1/2}m^{-5/2}\dot{m}^{2}R^{-15/4},\\
&P_\mathrm{deg}/P_\mathrm{gas}=2.12\times10^{-2}A^{1/3}B^{1/4}C^{1/12}\mathscr{D}^{-2/3} \alpha^{1/6}m^{5/6}\dot{m}^{-2/3}R^{5/4}.
\end{align}
\label{eq:C17}
\end{subequations}

If there is significant non-zero torque at the inner edge, a new solution (region IV) exists inside region III. In this region IV (see the next subsection), the disk is too hot to be dominated by radiation pressure, and thus becomes unstable. The dividing line between region IV and region III is the position satisfying $P_\mathrm{rad}/P=0.2$. But when the boundary torque vanishes or be too small, there will be another solution (region VI) just inside region III, where the degeneracy pressure becomes comparable with the gas pressure. The dividing line between region VI and region III are given by $P_\mathrm{deg}/P_\mathrm{gas}=1$.

\subsection{Region IV --- unstable NDAF}
From the results of numerical solution, we found that an viscously unstable region will appear in the inner vicinity of the grey NDAF if the inner edge torque is large enough (e.g., see $\dot m=1$, $\eta=3$ compared with $\dot m=1$, $\eta=0$ in Figure \ref{Fig:C1} and \ref{Fig:C2}), since the temperature in this region is so efficiently increased due to the additional viscous heating due to the boundary torque that the radiation pressure dominates the neutrino cooling flow. The neutrino energy density, i.e., $u_{\nu}= (7/8)a T^4\sum(\tau_{\nu_i}/2+1/\sqrt{3})/(\tau_{\nu_i}/2+1/\sqrt{3}+1/3\tau_{a,\nu_i})$ is reduced to $21aT^4/8$. Thus, the neutrino pressure in this region can be simply written as to $P_\nu=u_\nu/3\approx7aT^4/8$, which is comparable to the radiation pressure. In this case, the equation of state $P\approx (1-\beta)^{-1}(P_\mathrm{rad}+P_\nu)$ and energy conservation $Q_\mathrm{vis}^+\approx Q_\nu^-$ are expressed as

\begin{subequations}
\begin{align}
&\left(\frac{GM\dot{M}\rho^{1/2}}{4\pi \alpha r^3}\frac{C\mathscr{D}}{A^2}\right)^{2/3}\approx (1-\beta)^{-1} \frac{43}{24}aT^4,\\
&\frac{3GM\dot{M}}{8\pi r^{3}}\frac{\mathscr D}{B}\approx1.16\times10^{35}\rho^{-1}T^2h^{-1}\approx1.16\times10^{35}\rho^{-1}T^2 \left((1-\beta)\frac{24GM\rho}{43aT^4r^3}\frac{C}{B}\right)^{1/2}.
\end{align}
\label{eq:C18}
\end{subequations}

\noindent The solutions for this unstable NDAF are,

\begin{subequations}
\begin{align}
&\rho=4.27\times10^{8}(1-\beta)BC\mathscr{D}^{-2}m^{2}\dot{m}^{-2}R^{3} \mathrm{\ \ \ g\ cm^{-3}},\\
&T=2.10\times10^{11} (1-\beta)^{1/3} A^{-1/3}B^{1/12}C^{1/4}\alpha^{-1/6}m^{-1/6}R^{-1/4} \mathrm{\ \ \ K},\\
%&h=1.22\times10^{6}A^{-2/3}B^{1/6}C^{-1/2}\mathscr{D}\alpha^{-1/3}m^{-1/3}\dot{m}R^{-1/2} \mathrm{\ \ \ cm},\\
&H=8.28(1-\beta)^{-1/3}A^{-2/3}B^{1/6}C^{-1/2}\mathscr{D}\alpha^{-1/3}m^{-4/3}\dot{m}R^{-1/2},\\
&u^r=2.05\times10^{12}(1-\beta)^{-2/3}A^{2/3}B^{-7/6}C^{-1/2}\mathscr{D}\alpha^{1/3}m^{-8/3}\dot{m}^{2}R^{-7/2} \mathrm{\ \ \ cm\ s^{-1}},\\
&Q_{\nu}^-=Q_{\rm{vis}}^+=9.79\times10^{42} (1-\beta)^{2/3} B^{-1}\mathscr{D}m^{-2}\dot{m}R^{-3} \mathrm{\ \ \ erg\ cm^{-2}\ s^{-1}},\\
&\Sigma=1.04\times10^{15}(1-\beta)^{2/3}A^{-2/3}B^{7/6}C^{1/2}\mathscr{D}^{-1} \alpha^{-1/3}m^{5/3}\dot{m}^{-1}R^{5/2} \mathrm{\ \ \ g\ cm^{-2}},\\
&\partial\dot{M}/\partial\Sigma=-1.91\times10^{18}(1-\beta)^{-2/3}A^{2/3}B^{-7/6}C^{-1/2}\mathscr{D} \alpha^{1/3}m^{-5/3}\dot{m}^{2}R^{-5/2} \mathrm{\ \ \ cm^{2}\ s^{-1}},\\
&t_\mathrm{vis}=7.19\times10^{-8}(1-\beta)^{2/3}A^{4/3}B^{-1/3}C\mathscr{D}^{-2}\alpha^{-1/3}m^{11/3}\dot{m}^{-2}R^{9/2} \mathrm{\ \ \ s},\\
&P_\mathrm{gas}/P=2.81\times10^{-4}(1-\beta)AB^{3/4}C^{1/4}\mathscr{D}^{-2}\alpha^{1/2}m^{5/2}\dot{m}^{-2}R^{15/4},\\
&f\equiv Q_\mathrm{adv}^-/Q_\mathrm{vis}^+=1.23\times10^{2}(1-\beta)^{-2/3} A^{-4/3}B^{1/3}\mathscr{D} \alpha^{-2/3}m^{-8/3}\dot{m}^{2}R^{-3}.
\end{align}
\label{eq:C19}
\end{subequations}

We find that $\Sigma\propto\dot{m}^{-1}$ and $\partial\dot{M}/\partial\Sigma<0$. Therefore, this region is viscously unstable. The timescale of the instability $t_{\rm vis}$ in equation (\ref{eq:C19}h) has been evaluated as the viscous timescale, i.e., $t_\mathrm{vis}\sim r^2/\nu=(\alpha\Omega_\text{k})^{-1}(h/r)^{-2}$. For $\eta=3$ and $r=3 r_g$, $t_\mathrm{vis} \simeq 6.2\ \mathrm{ms}$ and $24.7\ \mathrm{ms}$ for $\dot m=0.5$, which are consistent with the numerical results in section 5.

Inside this unstable region, there is a new solution (region V) where the advection cooling dominates over radation cooling. The solution transits from IV to V at the location satisfying $Q_\mathrm{adv}^-/Q_\mathrm{vis}^+=0.9$.

\subsection{Region V --- radiation and neutrino pressures dominated ADAF}
Inside region IV, there is a region where the neutrino optical depth is so high that the neutrinos are trapped, resulting in an advection cooling dominated flow. The radiation and neutrino pressures are still dominated due to the high temperature and large neutrino opacity. The analytical solutions are nearly same as those in region I, excepting the the discrepancy in the coefficients due to the consideration of neutrino pressure.

\begin{subequations}
\begin{align}
&\rho=5.84\times10^{11} (1-\beta)^{3/2} A^{-2}B^{3/2}C\mathscr{D}^{-1/2}\alpha^{-1}m^{-2}\dot{m} R^{-3/2} \mathrm{\ \ \ g\ cm^{-3}}\\
&T=3.83\times10^{11}(1-\beta)^{3/8} A^{-1/2}B^{1/8}C^{1/4}\mathscr{D}^{1/8} \alpha^{-1/4}m^{-1/2}\dot{m}^{1/4}R^{-5/8} \mathrm{\ \ \ K}\\
%&h=1.10\times10^5C^{-1/2}\mathscr{D}^{1/2}mR \mathrm{\ \ \ cm}\\
&H=0.75 (1-\beta)^{-1/2} C^{-1/2}\mathscr{D}^{1/2}R\\
&u^r=1.67\times10^{10} (1-\beta)^{-1} A^2B^{-3/2}C^{-1/2}\alpha R^{-1/2} \mathrm{\ \ \ cm/s}\\
&f_\nu\equiv Q_\nu^-/Q_\mathrm{vis}^+=2.70\times10^{-2}(1-\beta)^{-1/4} AB^{-1/4}\mathscr{D}^{-3/4} \alpha^{1/2}m^2\dot{m}^{-3/2}R^{9/4}
\end{align}
\label{eq:C20}
\end{subequations}

This region is usually too narrow to be picked out (e.g. see the panels of Figure \ref{Fig:C1} and \ref{Fig:C2} in which $\dot{m}=1$, $\eta=3$ as an example, the region V starts from $r=2.6r_g$). To show this region V, we solve a disk with extremely large $\dot m$ and $\eta$, as illustrated in Figure \ref{Fig:C3}.

\subsection{Region VI --- gas and degeneracy pressure dominated ADAF}
Note that regions IV and V usually exist when the inner edge torque is great enough. For NDAF without bounary torque, these two regions will be replaced with another ADAF region dominated by degneracy pressure as well as gas pressure.In order to grasp the key property of such an innermost region analytically, we just keep the degeneracy pressure in equation of state, i.e., $P\approx (1-\beta)^{-1} P_\mathrm{deg}$. Whileas for the energy equation $Q_\mathrm{vis}^+\approx Q_\mathrm{adv}^-$, we note that the degeneracy pressure has no contribution to the advection term. So one has,

\begin{subequations}
\begin{align}
&\left(\frac{GM\dot{M}\rho^{1/2}}{4\pi \alpha r^3}\frac{C\mathscr{D}}{A^2}\right)^{2/3}\approx 4.89\times10^{14}(1-\beta)^{-1} \rho^{4/3},\\
&\frac{3GM\dot{M}}{8\pi r^{3}}\frac{\mathscr{D}}{B}\approx\frac{\dot M}{4\pi r^2 \rho} \frac{3}{4}\frac{\rho k T}{m_\mathrm{p}}.
\end{align}
\label{eq:C21}
\end{subequations}

\noindent The solutions of this region are,

\begin{subequations}
\begin{align}
&\rho=7.15\times10^{13}(1-\beta)A^{-4/3}C^{2/3}\mathscr{D}^{2/3}\alpha^{-2/3}m^{-4/3}\dot{m}^{2/3} R^{-2} \mathrm{\ \ \ g\ cm^{-3}}\\
&T=2.18\times10^{13}B^{-1}\mathscr{D}R^{-1} \mathrm{\ \ \ K}\\
%&h=2.09\times10^5C^{-1/2}\mathscr{D}^{1/2}mR \mathrm{\ \ \ cm}\\
&H=0.15(1-\beta)^{-1/3}A^{-2/9}B^{1/2}C^{-7/18}\mathscr{D}^{1/9}\alpha^{-1/9}m^{-2/9}\dot{m}^{1/9} R^{7/6}\\
&u^r=6.76\times10^{8}(1-\beta)^{-2/3}A^{14/9}B^{-1/2}C^{-5/18}\mathscr{D}^{-7/9}\alpha^{7/9} m^{-4/9}\dot{m}^{2/9} R^{-1/6} \mathrm{\ \ \ cm/s}
\end{align}
\label{eq:C22}
\end{subequations}

From Figure \ref{Fig:C4}, we find that the width of region VI is close to zero even at extremely large accretion rate, so this region is hard to exhibit in most cases.

\subsection{Effects of the inner edge torque}
\begin{figure}[!ht]
\centering
\includegraphics[width=0.9\textwidth]{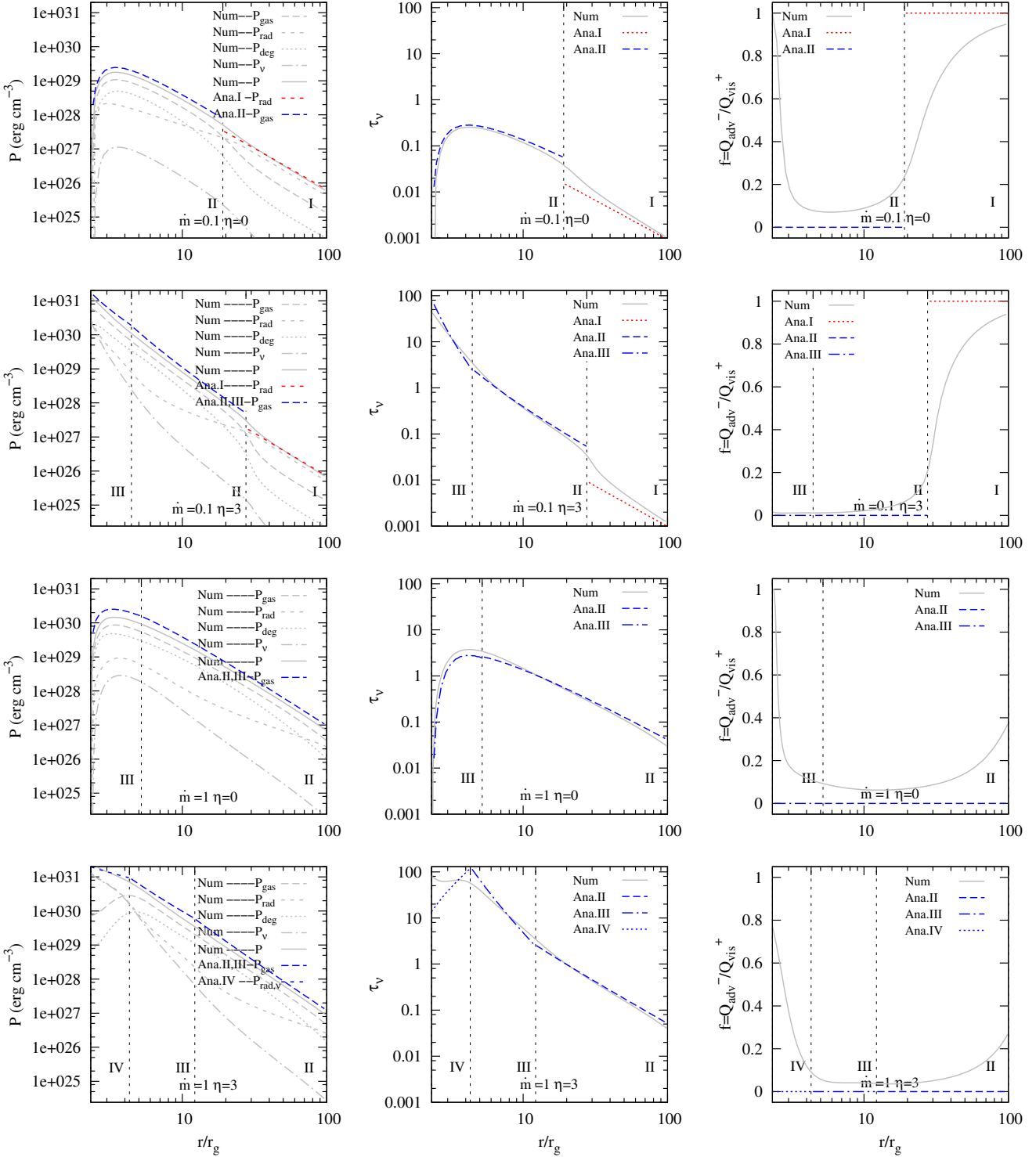}
\caption{The radial distribution of pressure, neutrino optical depth, and the cooling rate for different mass accretion rate ($\dot m=0.1,\ 1$) and inner edge torque ($\eta=0,\ 3$), the rest parameters are $m=7$, $a_*=0.9$, $\alpha=0.1$, $\beta=0$. The numerical results and the analytical results are separatively denoted by gray lines and black lines. The vertical dashed lines denote the transition between two neighbouring regions. The inner radiation pressure dominated ADAF (region V) can appear only if the unstable NDAF region exists in the disk due to a high inner edge torque, which can be certified from the bottom right graphic above, in which the region V starts from $r=2.6r_g$, it's too close to $r_\mathrm{ms}$ to be exhibited in the figure. Similarly, the width of region VI existing in the first and third row is nearly zero so that our analytical solution can hardly capture it. For a much clearer recognition of region V and VI, one can refer to Figrue \ref{Fig:C3} and \ref{Fig:C4}, corresponding to much extremmer parameters.}
\label{Fig:C1}
\end{figure}

\begin{figure}[!ht]
\centering
\includegraphics[width=0.9\textwidth]{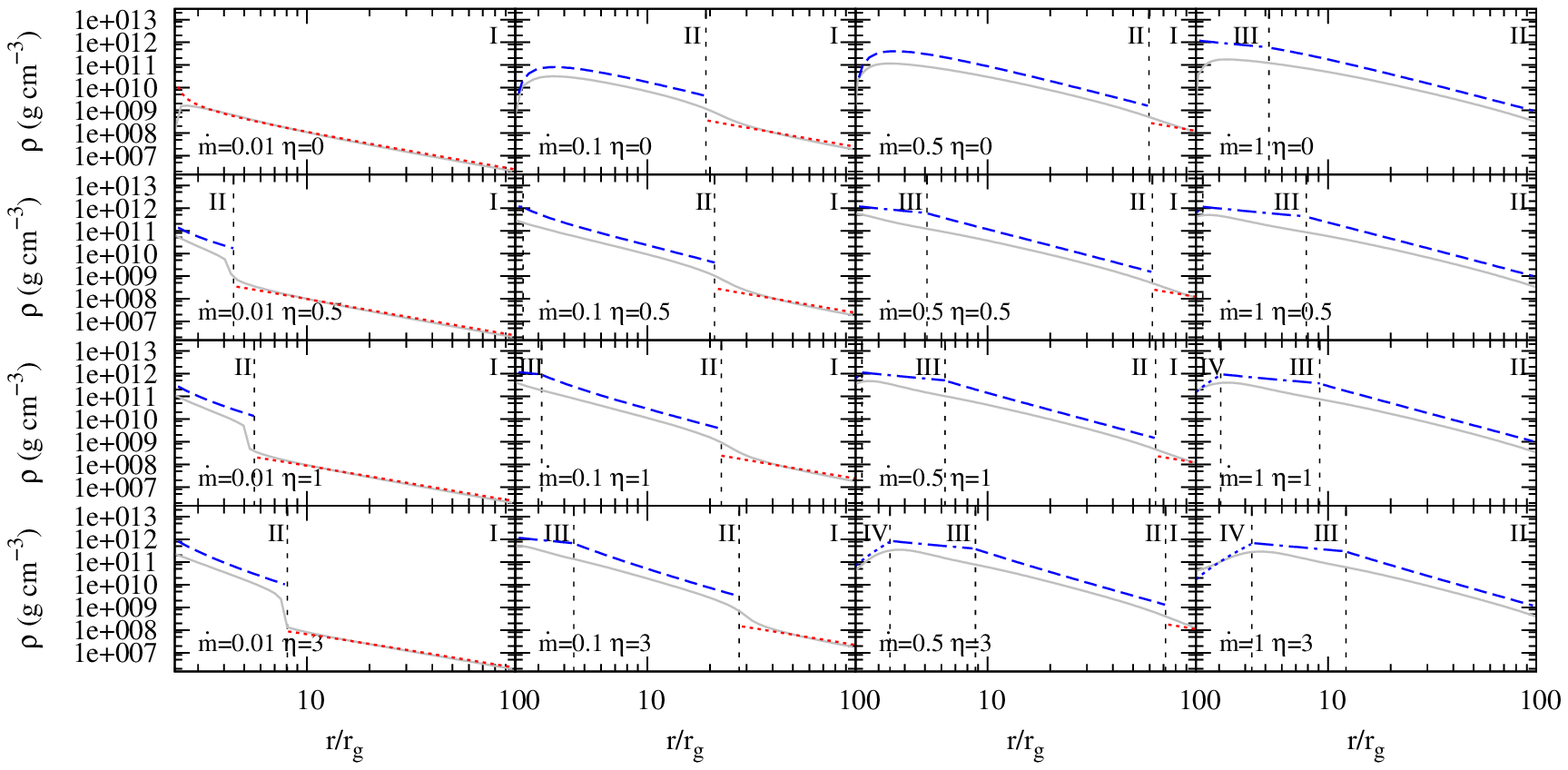}\\
\includegraphics[width=0.9\textwidth]{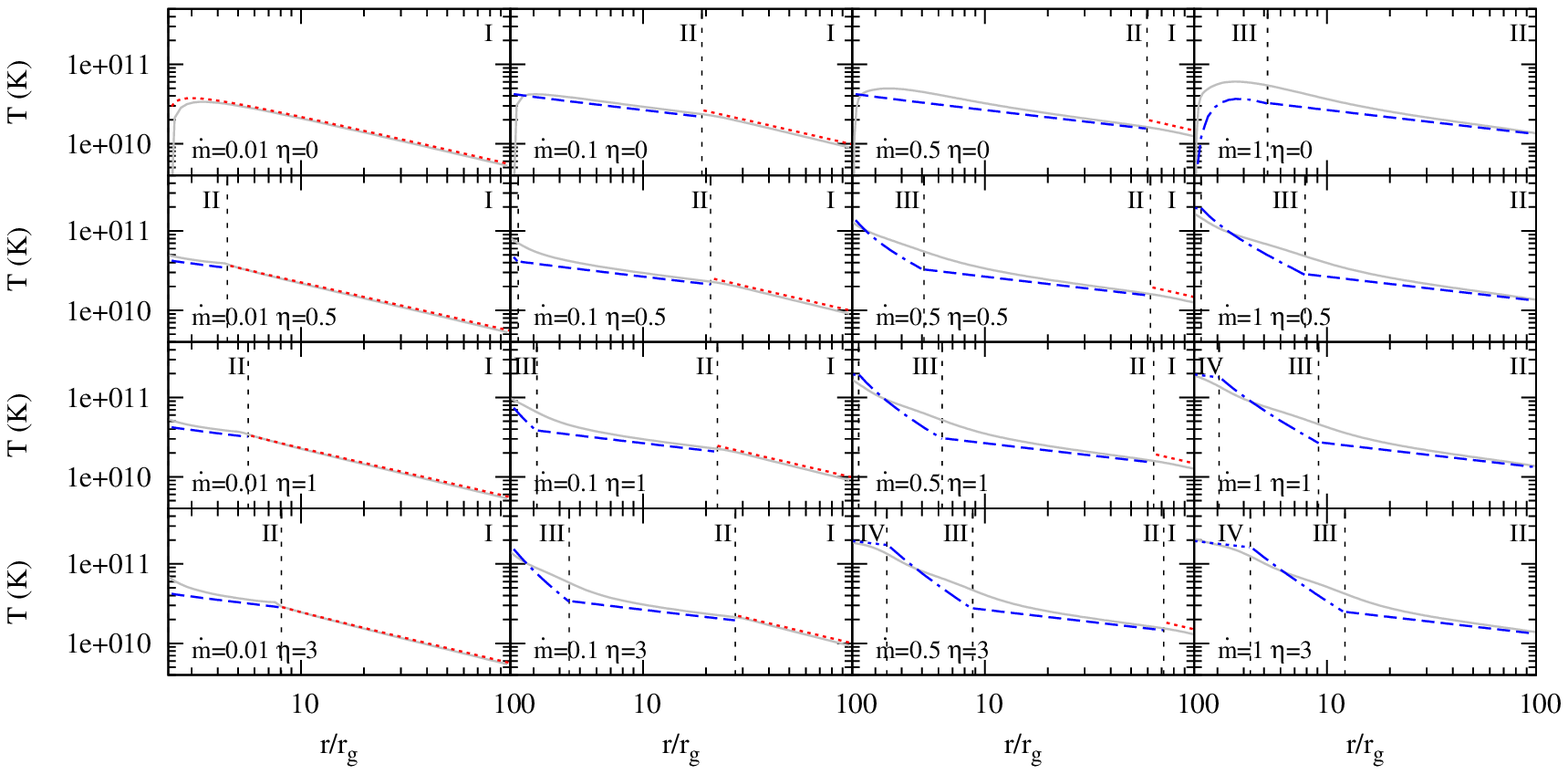}
\caption{The radial distribution of density, temperature, scale height and radial velocity for different parameters, i.e., $\dot m=0.01,\ 0.1,\ 0.5$ and $1$, $\eta=0,\ 0.5,\ 1$ and $3$, the rest parameters are $m=7$, $a_*=0.9$, $\alpha=0.1$, $\beta=0$. The numerical solutions and analytical solutions are separatively denoted by gray lines and colored lines. The vertical dashed lines denotes the transition between two neighbouring regions. The solutions of region I (radiation dominated ADAF) are denoted by red dotted lines, the solutions of region II (transparent NDAF) are denoted by blue dashed lines, the solutions of region III (opaque NDAF) are denoted by blue dashdotted lines, the solutions of region IV (unstable NDAF) are denoted by blue dotted lines.}
\label{Fig:C2}
\end{figure}

From Figures \ref{Fig:C1} and \ref{Fig:C2}, we find that our analytical solutions are identical with the numerical solutions on the whole. Now we can summarize the effects of inner edge torque:

Firstly, the existence of the inner edge torque can reduce the ignition accretion rate, as shown in the first column of Figure \ref{Fig:C2}. At relative lower accretion rate $\dot m =0.01$, the entire disk is ADAF when there is no inner edge torque. However, once the inner edge torque increases to $\eta=0.5$, a NDAF region emerges in the inner disk. Furthermore, the NDAF region expands outward accompany with the increasing of $\eta$, since that the increase of $\eta$ ($\mathscr{D}$) leads to a larger mass density (see equation (\ref{eq:C13}a)) or a higher temperature (see equation (\ref{eq:C17}b)).

Secondly, the inner edge torque can enhance the neutrino luminosity effectively. This conclusion can be found by examining equations (\ref{eq:C13}e), (\ref{eq:C17}e), and (\ref{eq:C19}e).

Thirdly, the inner edge torque can cause the NDAF to be viscously unstable. As shown in the fourth column of Figure \ref{Fig:C2}, for $\dot m=1$, when the inner edge torque $\eta$ increases to $\eta=0.5$, an unstable region starts to emerge in the inner disk if $\eta > 0.5$ and expands with the increase of $\eta$.
\clearpage

\renewcommand{\thefigure}{\arabic{figure} (Cont.)}
\addtocounter{figure}{-1}
\begin{figure}[!ht]
%\ContinuedFloat
%\captionsetup{list=off,format=cont}
\centering
\includegraphics[width=0.9\textwidth]{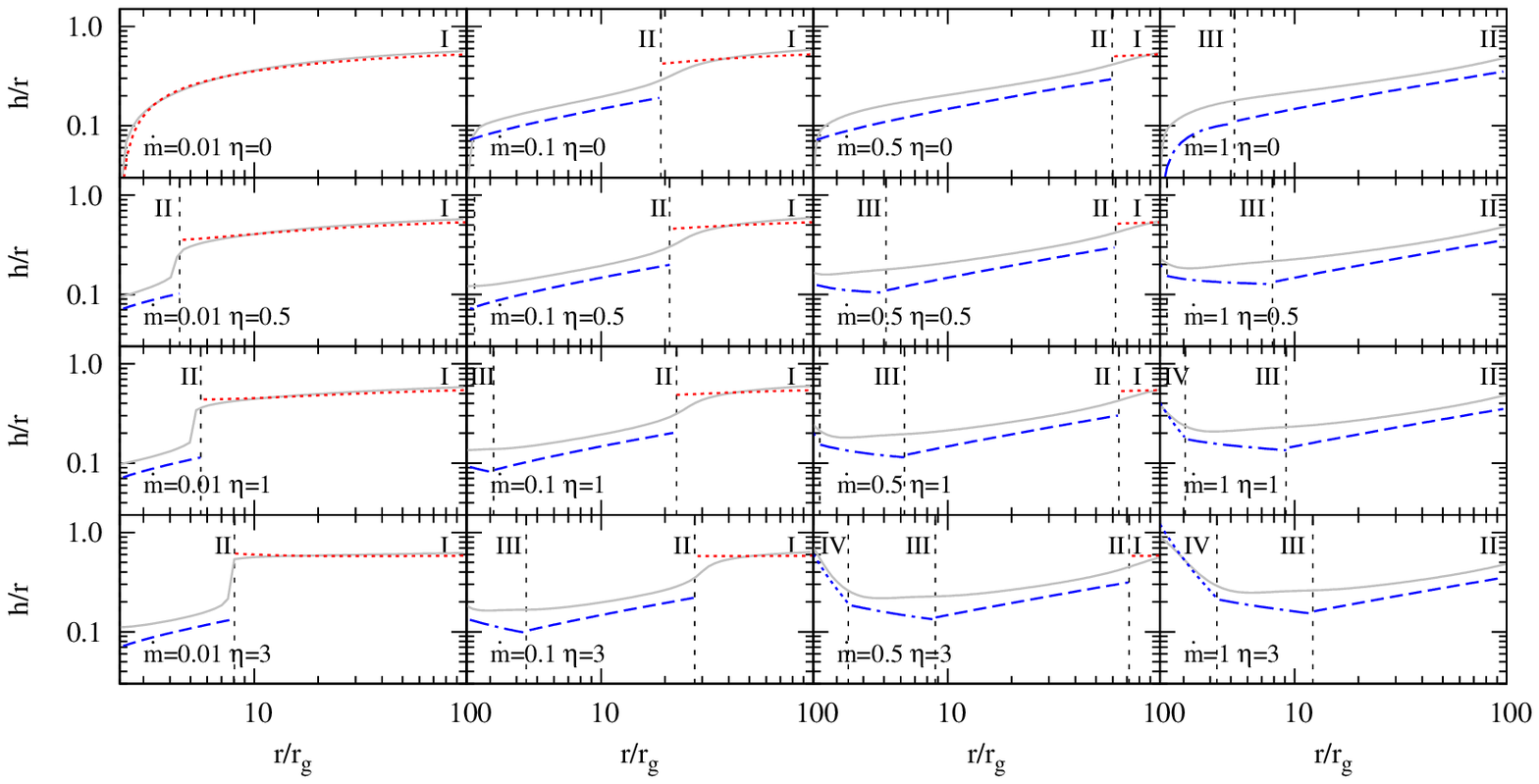}\\
\includegraphics[width=0.9\textwidth]{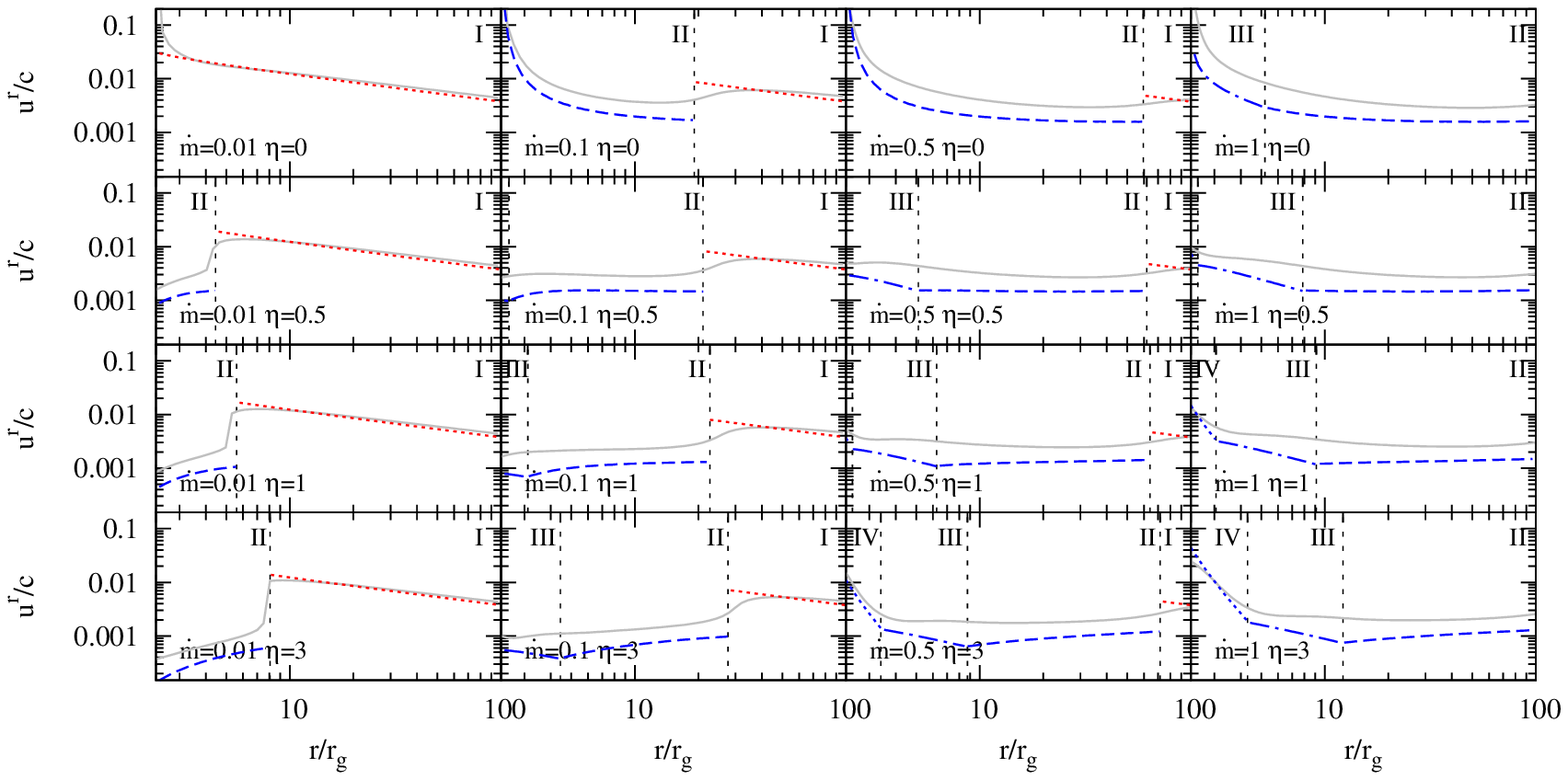}
\caption{The radial distribution of density, temperature, scale height and radial velocity for different parameters, i.e., $\dot m=0.01,\ 0.1,\ 0.5$ and $1$, $\eta=0,\ 0.5,\ 1$ and $3$, the rest parameters are $m=7$, $a_*=0.9$, $\alpha=0.1$, $\beta=0$. The numerical solutions and analytical solutions are separatively denoted by gray lines and colored lines. The vertical dashed lines denotes the transition between two neighbouring regions. The solutions of region I (radiation dominated ADAF) are denoted by red dotted lines, the solutions of region II (transparent NDAF) are denoted by blue dashed lines, the solutions of region III (opaque NDAF) are denoted by blue dashdotted lines, the solutions of region IV (unstable NDAF) are denoted by blue dotted lines.}
\end{figure}
\renewcommand{\thefigure}{\arabic{figure}}

\begin{figure}[!ht]
\centering
\includegraphics[width=0.8\textwidth]{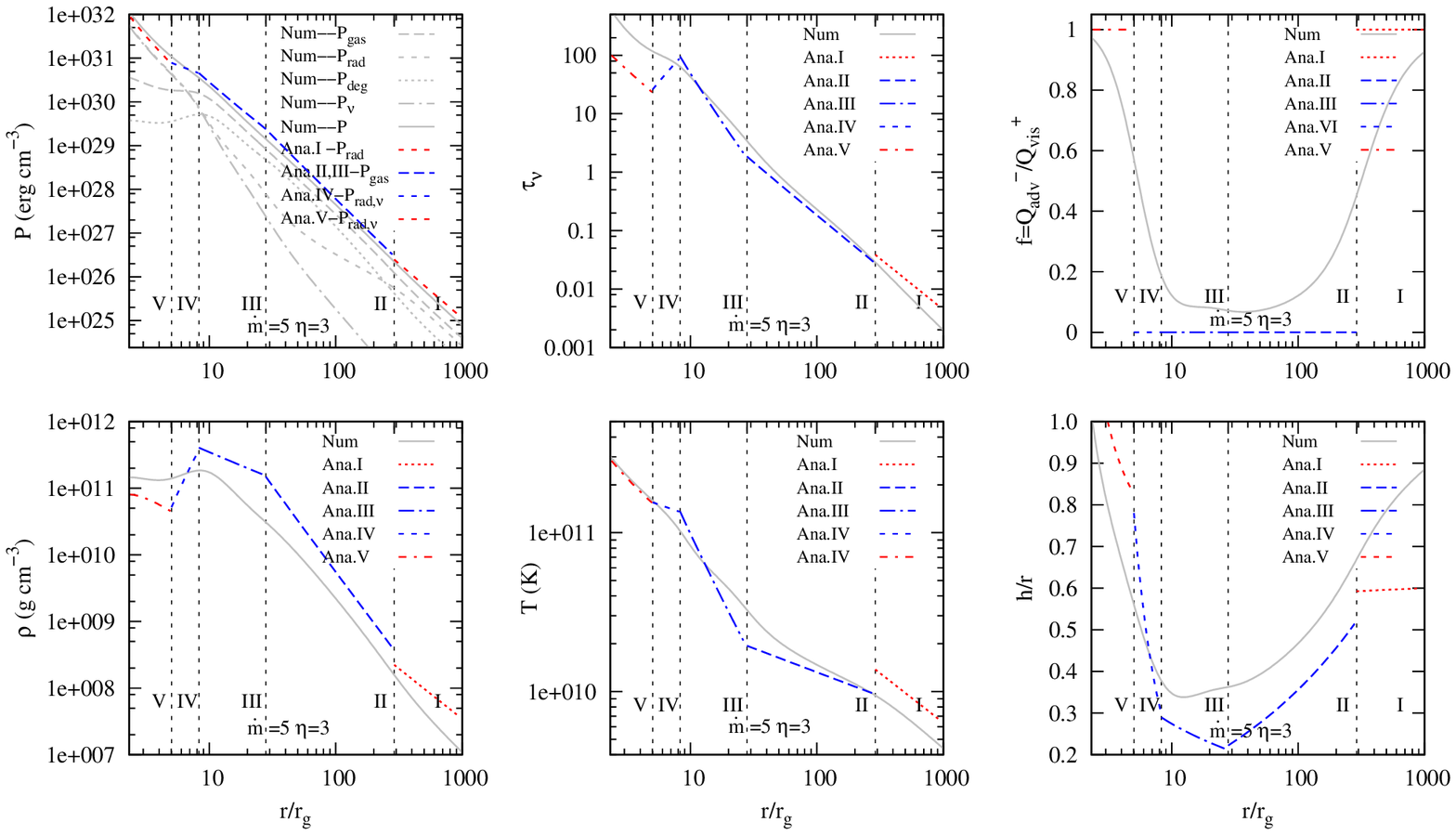}
\caption{Same as Figure \ref{Fig:C1} and \ref{Fig:C2}, but focused on the existence of region V starting from $r=5.0r_g$, other parameters are $\dot m=5$, $\eta=3$, $\beta=0$, etc. This scenario stands for the characteristic structure of nztNDAF as shown in the right panel of Figure \ref{sketch}.}
\label{Fig:C3}
\end{figure}

\begin{figure}[!ht]
\centering
\includegraphics[width=0.8\textwidth]{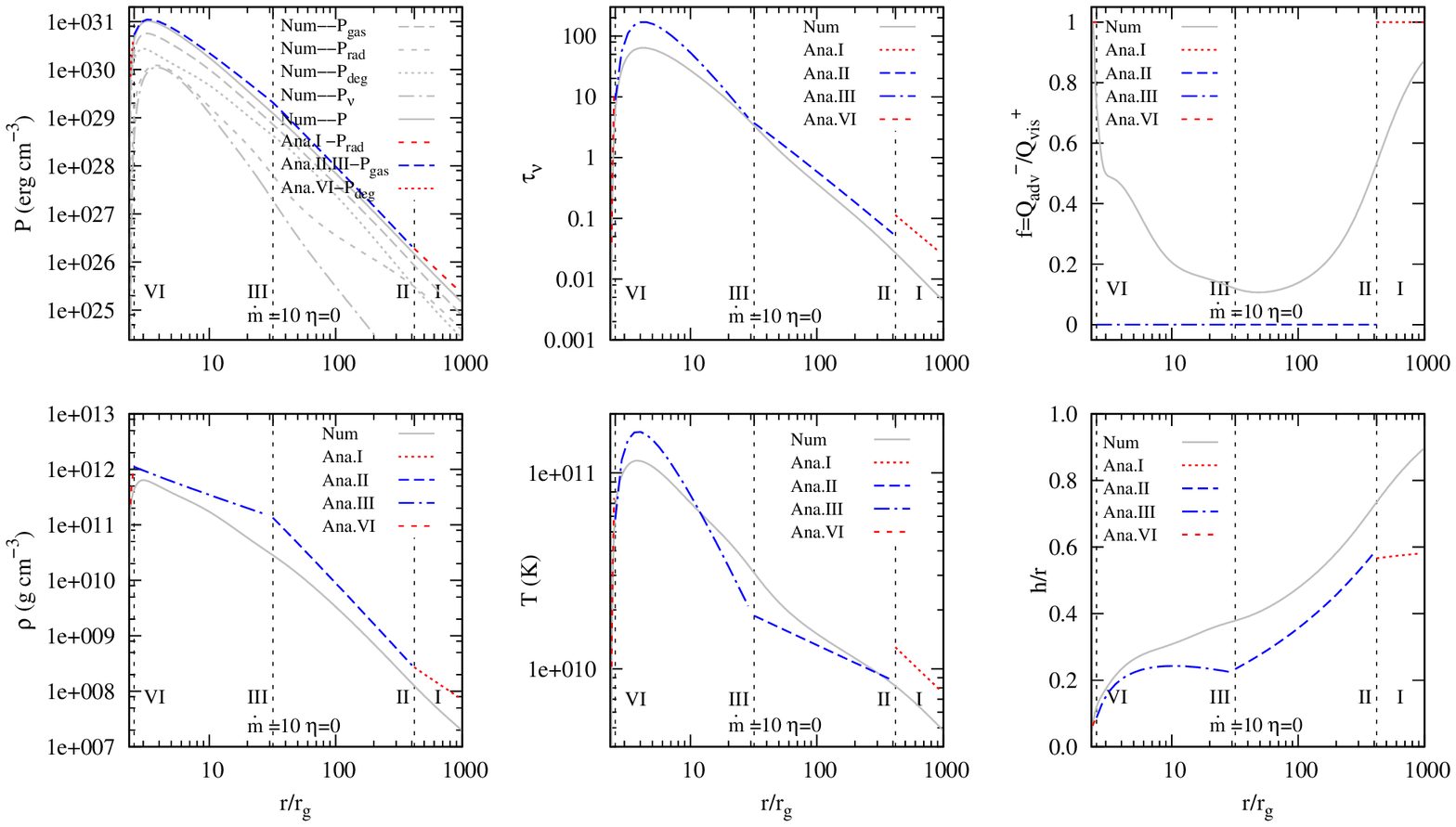}
\caption{Same as Figure \ref{Fig:C1} to \ref{Fig:C3}, but focused on the existence of region VI starting from $r=2.6r_g$, other parameters are $\dot m=10$, $\eta=0$, $\beta=0$, etc. This scenario stands for the characteristic structure of NDAF as shown in the left panel of Figure \ref{sketch}.}
\label{Fig:C4}
\end{figure}

\end{document}